\documentclass[a4paper,11pt]{article}
\usepackage{jcappub} 

\usepackage[T1]{fontenc}
\usepackage{booktabs}
\usepackage{graphicx} 
\usepackage{caption}
\usepackage{subcaption} 
\usepackage{lipsum}
\usepackage{xcolor}
\usepackage{float}
\usepackage{amsmath}	
\usepackage{titlesec}
\usepackage{lineno}

\title{Initial Conditions from Galaxies: Machine-Learning Subgrid Correction to Standard Reconstruction}

\author[a,b]{Liam Parker,}
\author[c,d]{Adrian E.~Bayer,}
\author[a,b,e]{Uroš Seljak}
\affiliation[a]{Department of Physics, University of California, Berkeley,\\
366 LeConte Hall, Berkeley, CA 94720, U.S.A.}
\affiliation[b]{Berkeley Center for Cosmological Physics, University of California, Berkeley,\\
341 Campbell Hall, Berkeley, CA 94720, U.S.A.}
\affiliation[c]{Department of Astrophysical Sciences, Princeton University,\\
Peyton Hall, Princeton, NJ 08544, U.S.A.}
\affiliation[d]{Center for Computational Astrophysics, Flatiron Institute,\\
162 5th Avenue, New York, NY 10010, U.S.A.}
\affiliation[e]{Physics Division, Lawrence Berkeley National Laboratory,\\
1 Cyclotron Road, Berkeley, CA 94720, U.S.A.}

\emailAdd{lhparker@berkeley.edu}

\abstract{
We present a hybrid method for reconstructing the primordial density from late-time halos and galaxies.
Our approach involves two steps: (1) apply standard Baryon Acoustic Oscillation (BAO) reconstruction to recover the large-scale features in the primordial density field and (2) train a deep learning model to learn small-scale corrections on partitioned subgrids of the full volume. At inference, this correction is then convolved across the full survey volume, enabling scaling to large survey volumes. 
We train our method on both mock halo catalogs and mock galaxy catalogs in both configuration and redshift space from the \textsc{Quijote} \(1(h^{-1}\,\mathrm{Gpc})^3\) simulation suite. When evaluated on held-out simulations, our combined approach significantly improves the reconstruction cross-correlation coefficient with the true initial density field and remains robust to moderate model misspecification. 
Additionally, we show that models trained on $1(h^{-1}\,\mathrm{Gpc})^3$ can be applied to larger boxes—e.g., $(3h^{-1}\,\mathrm{Gpc})^3$—without retraining. Finally, we perform a Fisher analysis on our method's recovery of the BAO peak, and find that it significantly improves the error on the acoustic scale relative to standard BAO reconstruction. Ultimately, this method robustly captures nonlinearities and bias without sacrificing large-scale accuracy, and its flexibility to handle arbitrarily large volumes without escalating computational requirements makes it especially promising for large-volume surveys like DESI.
}

\begin{document}
\maketitle
\flushbottom

\section{Introduction}
\label{sec:intro}

In recent decades, rapid advances in galaxy surveys have mapped ever-larger volumes of the universe at increasingly higher precision. These surveys reveal the rich clustering of matter, which encodes information on the universe's expansion history, its growth of structure, and a variety of cosmological parameters that govern its dynamics. Ongoing surveys like the Dark Energy Spectroscopic Instrument \citep[DESI,][]{aghamousa2016desi} and upcoming projects like Euclid \citep{laureijs2011euclid}, the Vera C. Rubin Observatory \citep{Ivezic2009}, and the Roman Space Telescope \cite{eifler2021cosmology} will push these insights even further.

A long-standing goal in cosmology is to reconstruct the primordial density field from the late-time observations, which have been nonlinearly evolved by gravity. By undoing the effects of gravitational clustering, one can more directly access early-universe signatures, such as baryon acoustic oscillations (BAO) or primordial non-Gaussianity. These are otherwise partially lost or blurred by nonlinear evolution in the late-time fields \citep{meiksin1999baryonic, eisenstein2007improving, crocce2008nonlinear}, due to the fact that late-time structure growth damps and broadens the BAO peak. Reconstructing the primordial density field reverses these non-linearities, sharpening the BAO. Moreover, it can potentially expose primordial non-Gaussianity that would otherwise remain hidden. Finally, reconstruction can also benefit full-shape analyses of cosmological parameters \citep{white2015reconstruction, chen2020consistent, floss2024improving}.

\cite{eisenstein2007improving} first demonstrated that the BAO feature can be sharpened by reconstructing the primordial field using first-order Lagragian perturbation theory; this method, dubbed ``standard reconstruction'', has been subsequently refined using iterative processes, annealing smoothing scales, and second-order perturbation theory \citep[e.g.,][]{seo2010high, padmanabhan20122, seo2016foreground, schmittfull2017iterative, hada2018iterative}. Ultimately, standard reconstruction based methods have proven to be highly effective in recovering large-scale modes near the BAO scale $\sim 150\, h^{-1}\,\mathrm{Mpc}$ and have become the prevailing reconstruction method for improving BAO measurements in galaxy surveys \citep{anderson2014clustering, alam2017clustering}. 

However, these methods rely largely on the Zel’dovich approximation \citep{zel1970gravitational}, which assumes that displacements are governed primarily by large-scale, irrotational flows. While sufficient to sharpen the BAO feature, this linear assumption breaks down at smaller, nonlinear scales where shell-crossing, virialization, and higher-order velocity fields become important \citep{valageas2011impact}. Moreover, galaxies and halos do not trace the underlying dark matter field uniformly: their bias can vary with scale, mass, or environment, further complicating efforts to fully “rewind” the field in highly clustered regions \citep{mehta2011galaxy, yu2017halo}. Consequently, standard reconstruction algorithms struggle to accurately restore information beyond the BAO scale, especially when reconstructing from biased tracers.

An alternative approach pursued in recent years has been forward 
modeling with differentiable N-body simulations coupled with
hierarchical Bayesian analysis to reconstruct the entire linear field. Two main approaches in this regard are optimization towards Maximum A Posteriori solution (MAP) \cite{Seljak_2017, modi2018cosmological, Bayer:2022vid} or Monte Carlo Markov Chain sampling \cite{Jasche_2013, Jasche:2018oym, Ramanah:2018eed, Schmidt:2018bkr, Schmidt:2020viy, Nguyen:2020hxe, Kostic:2022vok, Bayer:2023rmj, Nguyen:2024yth}. The BAO can then be measured in this reconstructed linear field, with \cite{modi2018cosmological} showing that the MAP approach significantly outperforms standard BAO reconstruction, and \cite{Babic:2022dws, Babic:2024wph} showing improvements of up to $50\%$ on the BAO scale using Monte Carlo with an Effective Field Theory forward model. In principle this approach should be optimal, however, in practice, it is computationally expensive and potentially suffers from model
misspecification as the tracers are typically modeled with a bias model to enable fast computation and differentiability, which breaks down on small scales.

An alternative approach is to use deep learning. Deep learning methods have shown remarkable promise at recovering details of small-scale structure. These methods typically rely on simulation-based inference, where cosmological N-body simulations \citep[e.g.,][]{villaescusa2020quijote} generate late-time fields from a range of initial conditions. The models are then trained to learn the mapping from the late-time field to the initial conditions from these forward-modeled samples. For example, \cite{mao2021baryon} use a convolutional neural network (CNN) to map late-time dark matter fields to their linear counterparts. Building on this work, \cite{shallue2023reconstructing} and \cite{chen2023effective} combine standard reconstruction with a CNN, demonstrating an improvement in dark matter--to--dark matter reconstruction. Other approaches for reconstruction included using optimal transport \cite{Nikakhtar:2022cik} to map between the late-time and primordial field, and using diffusion models \cite{legin2024posterior} to generate probabilistic posterior samples of the primordial field.

So far, deep learning reconstruction has been limited to dark matter--to--dark matter mapping. However, real observational data necessarily involves biased tracers (i.e., galaxies or halos), which are more sparse than dark matter and exhibit scale-dependent bias. These additional complexities make the reconstruction task more challenging, as one must account for both shot noise and potential non-linearities in the bias model. Moreover, applying deep learning methods over the full multi-gigaparsec volumes of surveys like DESI introduces an additional layer of difficulty: computational scalability. A single cosmological box may contain billions of voxels, easily surpassing standard GPU memory limits and forcing such analyses to resort to heavy downsampling or coarser grids, thereby losing small-scale information.

In this work, we build on \cite{shallue2023reconstructing} by similarly combining a CNN with standard reconstruction, but we address two major challenges that arise in practice. First, we apply our method to both dark matter halos and galaxies directly, demonstrating that we can accurately infer highly non-linear scales from biased tracers of the late-time density fields instead of the dark matter field itself. Second, we introduce a sliding-window (sub-grid) training strategy to circumvent GPU memory constraints. Since large-scale modes are already recovered by standard reconstruction, the CNN only needs to correct small-scale features, making it feasible to train on manageable sub-volumes while preserving high-resolution information. This is ensured with a Fourier-space loss that explicitly ignores modes well described by standard reconstruction, focusing the network’s capacity on smaller scales where nonlinearities and shot noise dominate. Because each training sub-volume is processed independently, this sliding-window approach straightforwardly generalizes to arbitrary grid sizes, allowing the model to be applied to simulations of varying resolution and extent without retraining.

This paper is organized as follows. In \S\ref{sec:methods} we describe our reconstruction method, including our standard reconstruction algorithm in \S\ref{sec:standard-recon} and the learned subgrid correction's architecture, loss, and inference-time algorithm in \S\ref{sec:cnn}. We present our results in \S\ref{sec:results} for both dark matter halos \S\ref{sec:results-halos} and galaxies \S\ref{sec:galaxy_results}. For both, we present results in both configuration and redshift space, and for halos we present the effects of halo mass scatter. Additionally, we demonstrate how our method can be used on larger simulation volumes with no retraining in \S\ref{sec:scaling_analysis}, a crucial feature for upcoming galaxy surveys, as well as its robustness to a variety of model misspecifications in \S\ref{sec:model-mis}. Finally, we present a Fisher analysis of the recovery of the BAO peak in our reconstruction compared to standard reconstruction in \S\ref{sec:bao-info}. We conclude with a discussion on future directions and limitations.

\section{Methods}
\label{sec:methods}

Below, we present our two-step reconstruction procedure for inferring the primordial density field from biased tracers. First, an iterative FFT-based standard reconstruction estimates large-scale displacements, effectively “rewinding” the matter distribution near and above the BAO scale. Next, a learned correction via a convolutional neural network (CNN) recovers the finer-scale information lost to nonlinearities and bias. We train the CNN to reconstruct the primordial density field from both late-time halo fields and galaxy fields (in both real- and redshift-space) and the corresponding standard reconstructions using a sliding-window sub-grid approach that reduces memory demands without sacrificing resolution.

\subsection{Standard Reconstruction}
\label{sec:standard-recon}
The goal of standard reconstruction is to compute the Lagrangian displacement vector, $\boldsymbol{\Psi}(\textbf{q}, t)$, that maps a particle at some initial position $\textbf{q}$ to its Eulerian position at some later time $t$, 
\begin{align}
\textbf{x}(\textbf{q},t) = \textbf{q} + \boldsymbol{\Psi}(\textbf{q},t).
\end{align}
By ``undoing'' the computed displacements of particles in the late-time field
, one effectively ``rewinds'' the gravitational growth of structure, thereby recovering a representation that is closer to the initial conditions of the universe.

\subsubsection{Zel’dovich approximation} 
The Zel’dovich approximation \citep{zel1970gravitational} is frequently used to estimate $\boldsymbol{\Psi}$. It assumes the displacement field is proportional to the gradient of the gravitational potential, simplifying the relation between the overdensity and the displacement. In real space, $\boldsymbol{\Psi}$ satisfies
\begin{align}
    \nabla \cdot \boldsymbol{\Psi} = - \frac{\delta}{b},
\end{align}
where $\delta$ is the smoothed overdensity field and $b$ is the bias factor linking the observed tracer overdensity to the underlying matter overdensity.

In redshift space, the line-of-sight component $(\boldsymbol{\Psi} \cdot \hat r) \hat r$ introduces additional distortions proportional to the growth rate $f$. This adds a divergence term, modifying the relationship from the one above to
\begin{equation}
    \nabla \cdot \boldsymbol{\Psi} + f\nabla \cdot (\boldsymbol{\Psi} \dot \hat r) \hat r = - \frac{\delta}{b},
\label{eq:rsd_lag}
\end{equation}
where $\hat r$ represents the line-of-sight direction.

\subsubsection{Iterative FFT Displacement}
To solve for the displacement field, we use an iterative FFT-based reconstruction algorithm (\textsc{IterativeFFT}, \cite{burden2015reconstruction}), which is faster than configuration-space methods and provides accurate estimates of $\boldsymbol{\Psi}$ at moderate grid resolutions.

Like its configuration-space counterparts, \textsc{IterativeFFT} assumes $\boldsymbol{\Psi}$ is irrotational. However, because the redshift space term $(\boldsymbol{\Psi} \cdot \hat r) \hat r$ is not irrotational, Eq.~\ref{eq:rsd_lag} cannot be easily solved in Fourier space. To address this, $(\boldsymbol{\Psi} \cdot \hat r) \hat r$ is decomposed into an irrotational component (the gradient of a scalar potential field) and a solenoidal component (the curl of a vector field). The amplitude of these components is estimated using a plane parallel approximation in Fourier space, $\mathbf{\hat{r}} \rightarrow  \mathbf{\hat{x}}$, of the redshift-space distortions. Applying this recursively yields the displacement field in Fourier space as
\begin{equation}
    \boldsymbol{\Psi}_{\textrm{FFT, n}} = - \frac{i \textbf{k} \delta_{\textrm{g, real}}}{bk^2} \left( 1 + (-f)^{n+1} (k_x/k)^{2(n+1)} \right),
\label{eq:ifft}
\end{equation}
where $\textrm{n}$ denotes the iteration number.

\subsubsection{Practical Steps}
We first smooth the tracer overdensity field with a Gaussian kernel of width $\sigma = 10\,h^{-1}\,\text{Mpc}$ to reduce small-scale nonlinearities. Next, we initialize $20*N$ random particles—twenty times the number of tracers—and distribute them uniformly, throughout the box.

We then solve Eq.~\eqref{eq:ifft} iteratively to obtain $\boldsymbol{\Psi}$ for both the tracers and the randoms. We approximate the tracer bias as
\begin{align}
    b = \textrm{argmin}_b \sum_{\mathbf{|k|} \in (0, 0.1)} (P_\mathrm{g}(\mathbf{k}) - b^2 P_\mathrm{l}(\mathbf{k}))^2,
\end{align}
where $P_\mathrm{g}(k)$ is the power spectrum of the tracer overdensity field after subtracting shot noise ($1/\bar n$) and $P_\mathrm{l}(k)$ is the power spectrum of the linear dark matter field. We note that this linear approximation to the bias is intrinsically flawed at small scales, as bias is known to vary with scale, mass, and environment \citep{mehta2011galaxy, yu2017halo} -- our CNN correction, described later, will account for these additional factors.

Once $\boldsymbol{\Psi}$ is generated, we shift both the tracers and the randoms by $-\boldsymbol{\Psi}(\textbf{x})$ and deposit them onto two separate grids using the Cloud-in-Cell mass-assignment scheme (CIC, \cite{hockney2021computer}). Our final reconstructed field is then 
\begin{align} 
\delta_\mathrm{rec}(\mathbf{x}) = \frac{\hat \delta_\mathrm{g}(\mathbf{x}) - \hat \delta_\mathrm{r}(\mathbf{x})}{b},
\end{align}
where $\hat \delta_\mathrm{g}(\mathbf{x})$ and $\hat \delta_\mathrm{r}(\mathbf{x})$ are the displaced tracers and displaced randoms assigned to their respective grids. In practice, we use as input dark matter halos or galaxies at $z=0.5$, corresponding to a growth factor of $f \approx 0.76$.

\subsection{Learned Subgrid Correction}
\label{sec:cnn}

Having obtained a preliminary estimate of the linear density field via standard reconstruction, we now refine the small-scale, nonlinear modes that it cannot fully recover. We implement this refinement using a convolutional neural network (CNN) trained in a sliding-window (sub-grid) fashion to handle large volumes without sacrificing resolution. We adopt a weighted Fourier-space loss to emphasize the modes most impacted by nonlinearities and shot noise, while downweighting the large scales already captured by the standard reconstruction.

\subsubsection{Architecture}
The correction model takes as input subgrids of both the late-time tracer field, $\delta_{\mathrm{g}}$, and the BAO reconstruction approximation of the linear field field, $\delta_{\mathrm{rec}}$, and outputs a neural-network ``corrected'' linear field field:
\begin{equation}
    f: (\delta_{\mathrm{g}}(\mathbf{x}), \delta_{\mathrm{rec}}(\mathbf{x})) \rightarrow \delta_{\mathrm{l}}^{\mathrm{pred}}(\mathbf{x})
\end{equation}
Specifically, the input to the model is a subgrid of size $N_{\mathrm{sub}}^3 \times 2$ grid, where $N_{\mathrm{sub}}$ represents the spatial dimensions of the subgrid for each field and the two channels correspond to $\delta_{\mathrm{g}}$ and $\delta_{\mathrm{rec}}$. We pass the input through a series of 9 double-convolutional layers with \textsc{ReLU} activation. Each layer comprises one zero-padded convolutional layer and one non-padded layer, reducing the spatial dimension from $N_{\mathrm{sub}}$ to $(N_{\mathrm{sub}}-18)^3$, which helps avoid boundary artifacts by discarding edge regions. We use 32 channels for the first three layers, 64 for the next three, and 128 for the last three, before downsampling to a single channel for the final output. 

\subsubsection{Objective Function} 
We train the CNN to match the true initial density field from the simulations by minimizing a weighted Fourier-space loss:
\begin{align}
    \mathcal{L}_{\mathrm{Fourier}} = \sum_{\textbf{k}} M(k) \left| f(\Tilde{\delta}_{\mathrm{g}}(\textbf{k}), \Tilde{\delta}_{\mathrm{rec}}(\textbf{k})) - \Tilde{\delta}_{\mathrm{l}}(\textbf{k}) \right|^2,
\end{align}
where $\tilde \delta$ is the fourier transform of each field, and $M(k)$ is a mode-dependent weight:

\begin{align}
M(k) =
\begin{cases} 
10, & \text{if } k \in \left[0.08, 0.5\right] h\text{Mpc}^{-1}\\
1, & \text{otherwise}.
\end{cases}
\end{align}
We boost modes in the range $0.08 \leq k \leq 0.50 h\textrm{Mpc}^{-1}$ to prioritize small-scale features that standard reconstruction fails to capture. The upper bound corresponds to a slightly conservative cut on the scales at which the \textsc{Quijote} simulations are no longer reliable \citep{villaescusa2020quijote}; these scales are still fit by the model, but less emphasized than those in the $k$-range specified above.

\subsubsection{Inference on the Full Volume}
At inference time, we apply the learned CNN to the entire $N_{\mathrm{full}}^3$ volume. Specifically, we extract overlapping patches of size $N_{\mathrm{sub}}^3$ with an overlap of 40 voxels between neighboring patches. This ensures that every voxel within the full volume is predicted multiple times, but within slightly different local contexts, thereby increasing the effective receptive field for any given voxel. From each extracted patch, we infer a $(N_{\mathrm{sub}} - 18)^3$ region of primordial field predictions. All predictions within overlapping regions are then averaged, effectively ensembling the network’s outputs and yielding smoother transitions between patches. Through this procedure, we can tile arbitrarily large volumes despite having trained on relatively small subgrids, without being impacted by super-sample effects \citep{Bayer:2022nws}.

\subsection{Dataset}
\label{sec:dataset}

We use data from the \textsc{Quijote} N-body suite \citep{villaescusa2020quijote} at the fiducial cosmology which closely matches the Planck 2018 results \citep{aghanim2020planck}. In particular, we use one hundred of the high resolution fiducial simulations, which initialize \(1024^3\) cold dark matter particles at \(z=127\) using second-order Lagrangian perturbation theory within a \(1\,(h^{-1}\,\textrm{Gpc})^3\) volume, and then evolves them forward in time with the TreePM GADGET-III code to some final redshift. In this analysis, we choose a final redshift of \(z=0.5\) to approximate the DESI luminous red galaxy and bright galaxy survey redshift range. We also use boxes with volume \(3^3\,(h^{-1}\,\textrm{Gpc})^3\) run identically to the fiducial Quijote simulations \cite{Scoggins:2025bvt}, as well as the fiducial resolution Quijote runs, to demonstrate our method's scalability; these are all run at a resolution of \(512^3\) cold dark matter particles.

\subsubsection{Dark Matter Halos}

Dark matter halos are identified via the Friends-of-Friends algorithm (FoF), which groups particles according to a fixed linking length. We rank the halos by mass and select the top \(k\) to achieve two target number densities, \(\bar{n} = 10^{-3}\,(h/\textrm{Mpc})^3\) and \(\bar{n} = 5\times10^{-4}\,(h/\textrm{Mpc})^3\). These densities roughly correspond to the expected abundances of DESI Emission Line Galaxies (ELGs) and Luminous Red Galaxies (LRGs), respectively.

For each simulation, we add both mass-weighted and number-weighted halos to a grid using the Cloud-in-Cell (CIC) assignment scheme in both configuration space, where halos are placed at their comoving coordinates, and redshift space, where the halos’ \(z\)-positions are further displaced by their line-of-sight velocities. Finally, we convert these halo grids into overdensity fields and mesh them onto $N_{\mathrm{full}} = 256^3$ grids. 

Each halo field is paired with the corresponding linear dark matter overdensity field, which are computed by rescaling the initial linear field used by Quijote at \(z=127\) to \(z=0.5\), yielding 100 pairs of the halo field and linear field at $z=0.5$. We then randomly split these 100 pairs into a 90/10 training/testing set, ensuring that distinct simulations are used for training versus evaluation.

\subsubsection{Galaxy Catalogs}

We generate galaxy catalogs for each simulation by populating the halos with a state-of-the-art Halo Occupation Distribution (HOD) framework. Specifically, following the HOD strategy used in \cite{hahn2023simbig}, we augment the standard HOD model \cite[Z07,][]{zheng2007galaxy} with additional biases; hereafter, we call this augmented model Z07AB. The base Z07 model determines the expected numbers of central and satellite galaxies in a halo of mass \(M\) according to 
\begin{align}
    \langle N_\mathrm{c}(M)\rangle \; &= \; \frac{1}{2}\,\Bigg[1 + \mathrm{erf}\!\Bigg(\frac{\log M - \log M_{\min}}{\sigma_{\log M}}\Bigg)\Bigg],\\[6pt]
    \langle N_\mathrm{s}(M)\rangle \; &= \; \langle N_\mathrm{c}(M)\rangle \;\biggl(\tfrac{M - M_0}{M_1}\biggr)^\alpha,
\end{align}
where \(\log M_{\min}\) is the characteristic mass scale for halos to host a central galaxy, \(\sigma_{\log M}\) is the scatter of halo mass at a fixed galaxy luminosity, \(\log M_0\) is the minimum halo mass for halos to host a satellite galaxy, \(\log M_1\) is the characteristic mass scale for halos to host a satellite galaxy, and \(\alpha\) is the power-law index for the mass dependence of satellite occupation.

While broadly successful, recent work \cite{zentner2019constraints, vakili2019galaxies, hadzhiyska2021galaxy} has demonstrated that galaxy occupation relies on halo properties beyond $M_\mathrm{h}$. Therefore, Z07AB includes biases to account for assembly, concentration, and velocity biases. This introduces four additional parameters to the above: $A_{\mathrm{bias}}$ is the assembly bias, $\eta_\mathrm{conc}$ is the concentration bias, $\eta_{\mathrm{\mathrm{cent}}}$ is the velocity bias of centrals, and $\eta_{\mathrm{sat}}$ is the velocity bias of satellites. By incorporating these parameters, Z07AB allows halos of the same mass to host different numbers of galaxies depending on their formation history, change the concentration of the satellite profile relative to the host halo, and adjust the velocities of both central and satellite galaxies with respect to the underlying dark matter.

For each of our 100 halo catalogs, we sample a set of five Z07 HOD parameters from uniform priors (summarized in Table~\ref{tab:hod_params}) and a set of four additional HOD parameters to populate the Z07AB model. The prior ranges for these parameters are chosen to match the HOD priors used in the \cite{hahn2023simbig} analysis. Each halo catalog is sampled five times, once for each distinct draw from the five-dimensional HOD parameter space, resulting in a total of 500 galaxy catalogs. Note that these draws are not strictly independent from one another since they are all based on the same set of underlying halo catalogs; however, each realization corresponds to a distinct combination of HOD parameters.

\begin{table}
    \centering
    \begin{tabular}{cccc}
    \hline\hline
    Parameter & Prior Range & Description \\
    \hline
    \(\log M_{\min}\)  & \( \mathcal{U}(12, 14)\)    & Halo mass scale for a halo to host central galaxy   \\
    \(\sigma_{\log M}\)& \( \mathcal{U}(0.1, 0.6)\)   & Halo mass scatter at a fixed luminosity \\
    \(\log M_0\)       & \( \mathcal{U}(13, 15)\)    & Threshold halo mass for satellite suppression \\
    \(\log M_1\)       & \( \mathcal{U}(13, 15)\)    & Halo mass scale for halo to host a satellite galaxy \\
    \(\alpha\)         & \( \mathcal{U}(0, 1.5)\)    & Power law for satellite occupation mass dependence  \\
    \midrule
    $A_{\mathrm{bias}}$ & $\mathcal{N}(0, 0.2)$ over $[-1, 1]$ & Assembly bias \\
    $\eta_{\mathrm{conc}}$ & $\mathcal{U}(0.2, 2.0)$ & Concentration bias for satellites \\
    $\eta_{\mathrm{cent}}$ & $\mathcal{U}(0.0, 0.7)$ & Velocity bias for centrals \\
    $\eta_{\mathrm{sat}}$ & $\mathcal{U}(0.2, 2.0)$ & Velocity bias for satellites \\
    \hline
    \end{tabular}
    \caption{\textbf{Summary of the Halo Occupation Distribution (HOD) parameters sampled to generate the galaxy fields}. For each halo mock, we randomly sample a set of five HOD parameters to populate the standard Z07 \cite{zheng2007galaxy}, and augment this with four additional HOD parameters to account for assembly, concentration, and velocity biases. The priors used in this work are the same as those in \cite{hahn2023simbig}.}
    \label{tab:hod_params}
\end{table}

As with the halos, we generate real- and redshift-space versions of each galaxy catalog by incorporating line-of-sight velocity offsets for redshift space. Additionally, we use the same number densities as those in the halo catalog, namely $\bar n = 10^{-3} (h/\textrm{Mpc})^3$ and $\bar n = 5 \times  10^{-4} (h/\textrm{Mpc})^3$, corresponding to DESI ELGs and LRGs respectively. These catalogs are then binned onto $N = 256^3$ voxel grids using the same CIC assignment scheme as with halos and converted into galaxy overdensity fields for subsequent analysis. These are once again randomly split using a 90/10 training/testing set to ensure that distinct simulations are used for training and evaluation. Notably, we use the same 90/10 split for both halos and galaxy catalogs.

\subsection{Model Optimization}
\label{sec:optimization}

We compute the standard-reconstruction fields for each \textsc{Quijote} simulation prior to training, avoiding the computational overhead of recalculating them on-the-fly. We then train the CNN correction using stochastic gradient descent with the Adam optimizer \citep{kingma2014adam}. At each training iteration, we randomly sample an \((N_{\mathrm{sub}}^3 = 50^3)\) subgrid of \(\delta_g\) and \(\delta_{\mathrm{rec}}\), generating a predicted \(\delta_{l}^{\mathrm{pred}}\) output of size \((32^3)\). We compute the loss (described in the previous section) between \(\delta_{l}^{\mathrm{pred}}\) and the corresponding subgrid of the true initial density field \(\delta_l^{\mathrm{true}}\). 

We set the initial learning rate to \(\lambda=10^{-4}\) and decay it to \(\lambda=10^{-8}\) over the course of training using a cosine schedule. The model is trained for 100 epochs; each epoch consists of 80 random subgrid samples per simulation in the training set, ensuring thorough coverage of the simulation volume while keeping memory requirements modest. Although we only have 90 training simulations (with 450 HOD realizations in the galaxy catalog setting), subgridding substantially increases the data volume: from a \(256^3\) grid partitioned into \(50^3\) subgrids, we can extract approximately 134 subgrid samples. While many of these subgrids are correlated (thus not strictly i.i.d.), they still provide enough diversity across the full dataset that we do not observe significant overfitting, even with datasets that are relatively limited in scale by machine learning standards.

\section{Results}
\label{sec:results}
We evaluate our reconstructed linear density fields using the transfer function and cross-correlation coefficients presented in \S\ref{sec:metrics}. First, we discuss how our method performs when inferring initial conditions from dark matter halos (\S\ref{sec:results-halos}). Specifically, in \S\ref{sec:real-space}, we assess performance in configuration space; in \S\ref{sec:rsd}, and we investigate the impact of redshift-space distortions; in \S\ref{sec:halo-scatter}, we explore the effects of halo mass scatter. Next, we demonstrate reconstruction performance from galaxies \S\ref{sec:galaxy_results} in both configuration space in the presence of redshift-space distortions. Finally, we demonstrate that our method can be applied directly to larger simulation volumes in \S\ref{sec:scaling_analysis} without the need to retrain the model at all.

\subsection{Transfer function and cross-correlation coefficient}
\label{sec:metrics}
Throughout our analysis, we quantify the fidelity of reconstruction relative to the corresponding ground-truth linear density field in the \textsc{Quijote} suite with two power-spectrum--based measures. First, the transfer function,
\begin{align}
T(\mathbf{k}) \;=\; \sqrt{{\frac{P_{\mathrm{pred, pred}}(\mathbf{k})}{P_{\mathrm{truth,truth}}(\mathbf{k})}}},
\end{align}
compares the amplitude of the reconstructed field to the ground-truth initial field via their cross-correlation. It is desirable for the transfer function to be flat, such that it does not depend on cosmological parameters. However, for BAO reconstruction, even this is not a requirement, since  we are not interested in the broad band power. We will nevertheless show the results here to highlight the differences between our method and the standard reconstruction method. 

The second and more important 
quantification of fidelity is the cross-correlation coefficient,
\begin{align}
r(\mathbf{k}) \;=\; \frac{P_{\mathrm{pred,truth}}(\mathbf{k})}{\sqrt{P_{\mathrm{pred,pred}}(\mathbf{k})\;P_{\mathrm{truth,truth}}(\mathbf{k})}},
\label{rcc}
\end{align}
which measures how well the reconstructed field tracks the phase of the true initial field. In both expressions, \(P_{A,B}(k)\) denotes the power spectrum between two density fields \(A\) and \(B\):
\begin{align}
P_{\mathrm{A,B}}(\mathbf{k}) = \bigl\langle \widetilde{\delta}_A(\mathbf{k})\,\widetilde{\delta}_B^*(\mathbf{k}) \bigr\rangle.
\end{align}
For broad-band power, ideal performance corresponds to \(T(k) = 1\) and \(r(k) = 1\) across all relevant \(k\)-modes, while for BAO reconstruction only the 
latter is required, as we show in \S\ref{sec:bao-info}. 

\subsection{Dark Matter Halos}
\label{sec:results-halos}

\subsubsection{Configuration Space Results}
\label{sec:real-space}

\begin{figure*}
    \centering
    \begin{subfigure}[t]{\textwidth}
        \centering
        \includegraphics[width=\textwidth]{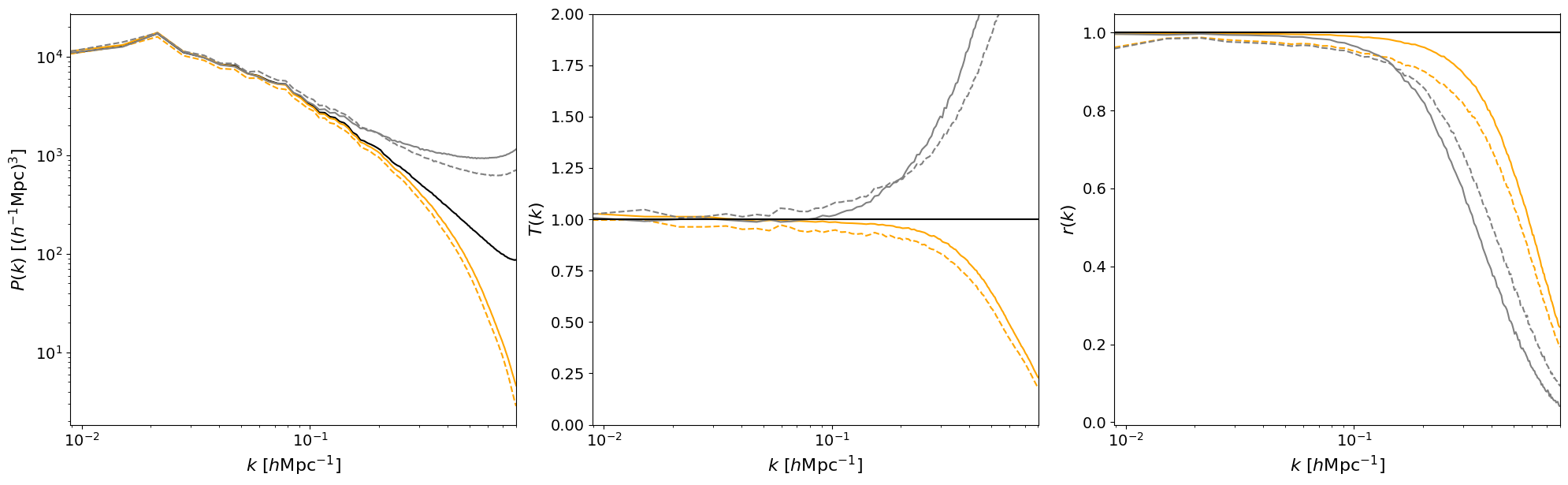}
        \label{fig:1e-3}
        \vspace{-0.5cm}
        \caption{$\bar n = 1 \times 10^{-3}$}
        \vspace{0.5cm} 
    \end{subfigure}
    \begin{subfigure}[t]{\textwidth}
        \centering
        \includegraphics[width=\textwidth]{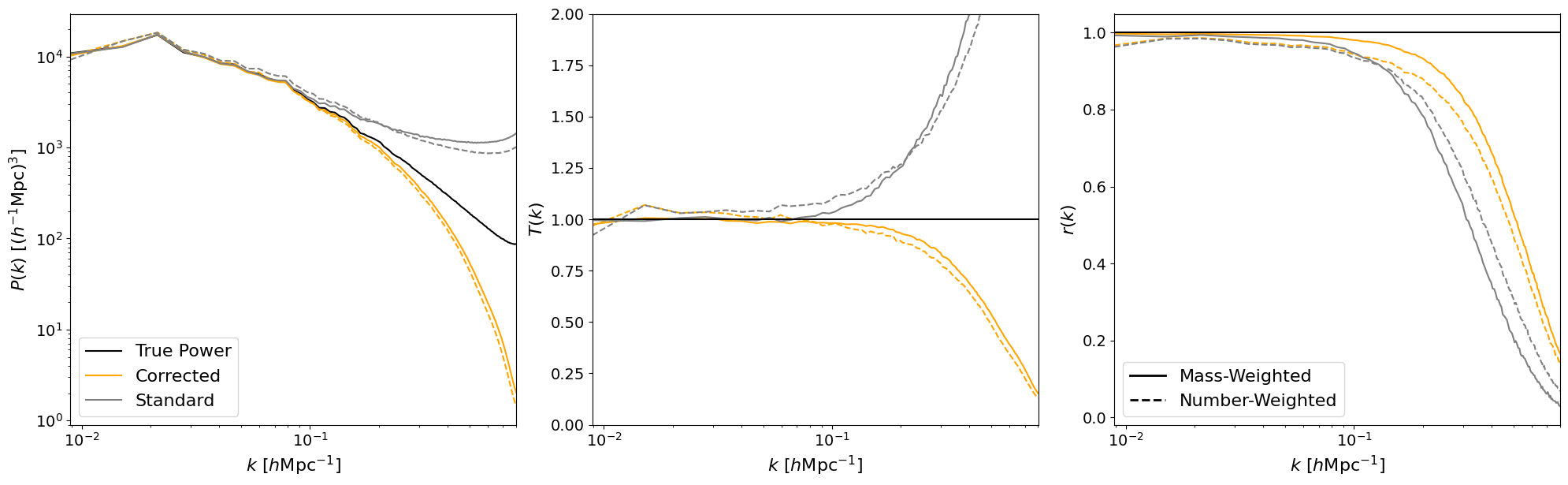}
        \label{fig:5e-4}
        \vspace{-0.5cm}
        \caption{$\bar n = 5 \times 10^{-4}$}
        \vspace{0.5cm}
    \end{subfigure}
    \caption{\textbf{Reconstruction of the linear dark matter field from late-time halos at \(\mathbf{z=0.5}\) in configuration space for two number densities.} Standard reconstruction is shown in gray, while our CNN-corrected approach is shown in orange. Solid lines indicate mass-weighted halos and dashed lines uniform number-weighting. The left panels compare the reconstructed power spectra to the true linear power (black), the middle panels show the transfer function \(T(k)\), and the right panels present the cross-correlation coefficient \(r(k)\). Standard reconstruction and our CNN correction converge on large scales, however our method consistently improves the recovered amplitude and correlation on intermediate to small scales where non-linearities dominate.}
    \label{fig:combined_reconstruction}
\end{figure*}

Figure~\ref{fig:combined_reconstruction} presents the performance of our CNN-corrected reconstruction (orange) versus standard reconstruction (gray) in configuration space for two halo number densities: \(\bar{n} = 1 \times 10^{-3}\) (top) and \(\bar{n} = 5 \times 10^{-4}\) (bottom), corresponding roughly to DESI ELG and LRG densities, respectively. In each panel, the solid lines correspond to reconstructions generated from a model trained on a mass-weighted halo field, while the dashed lines indicate those from a model trained on a uniform halo number-weighting. The left column compares the reconstructed power spectra to the true linear power spectrum at \(z=0.5\) (black), the middle column plots the transfer function \(T(k)\), and the right column shows the cross-correlation coefficient \(r(k)\). 

In the mass-weighted case (solid lines), the CNN-corrected prediction of the primordial density field more faithfully matches the true power at intermediate to small scales. In particular, the transfer function remains near unity to \(k \approx 0.2\text{--}0.3\,h\,\mathrm{Mpc}^{-1}\), while standard reconstruction (gray) begins to depart from the true amplitude at significantly larger scales. This improvement is also reflected in the cross-correlation coefficient, which exceeds \(r(k) \approx 90\%\) for modes as large as \(k \approx 0.3\,h\,\mathrm{Mpc}^{-1}\). Notably, the high-density sample \(\bigl(\bar{n} = 1\times10^{-3}\bigr)\) retains  better correlation on nonlinear scales compared to the sparser sample \(\bigl(\bar{n} = 5\times10^{-4}\bigr)\), but in both cases our approach significantly outperforms the standard method. For the number-weighted halo fields (dashed lines), we observe a similar trend, although small-scale modes become less correlated when halos are uniform weighted, due to the fact that uniform weighting of the halos does not reproduce the underlying mass distribution \cite{2010PhRvD..82d3515H,2013PhRvD..88h3507B}.

We note that a key component of our approach is that we train the CNN correction on sub-grids of the full simulation box; in this case, with size \(\sim 195\,h^{-1}\,\mathrm{Mpc}\) corresponding to \(50^3\) voxels. We then tile this trained CNN across the entire \(1(h^{-1}\,\mathrm{Gpc})^3\) volume at inference time, which is far more memory efficient than directly training on gigaparsec-scale grids. The excellent agreement with the true linear power in Fig.~\ref{fig:combined_reconstruction} (especially at large scales) demonstrates that using sub-grids in this setup does not degrade the recovery of long-wavelength modes. 

\subsubsection{Redshift-Space Distortions}
\label{sec:rsd}

As described in \S\ref{sec:dataset}, the \(z\)-positions of halos are displaced by their line-of-sight velocities, and standard reconstruction is performed on the velocity-shifted halo positions. We then retrain the CNN-correction on redshift-space halo fields before evaluating performance. 

We decompose the Fourier modes $\mathbf{k}$ by their magnitude $k = \|\mathbf{k}\|$ and by the cosine of their angle to a chosen line of sight (LOS). For a line-of-sight along the $z$-axis, this is
\[
    \mu \;=\; \frac{k_z}{\|\mathbf{k}\|}.
\]
We bin all modes $\mathbf{k}$ in discrete intervals of $k$ (ranging from $0$ up to the Nyquist frequency) and $\mu$ (centered on specified values). For each bin, we compute the cross-spectrum, the reconstructed auto-spectrum, and the ground-truth auto-spectrum. From the binned means, we compute the anisotropic cross-correlation coefficient \(r(k)\) as detailed above.

\begin{figure*}
    \centering
    \begin{subfigure}[b]{\textwidth}
        \centering
        \includegraphics[width=\textwidth]{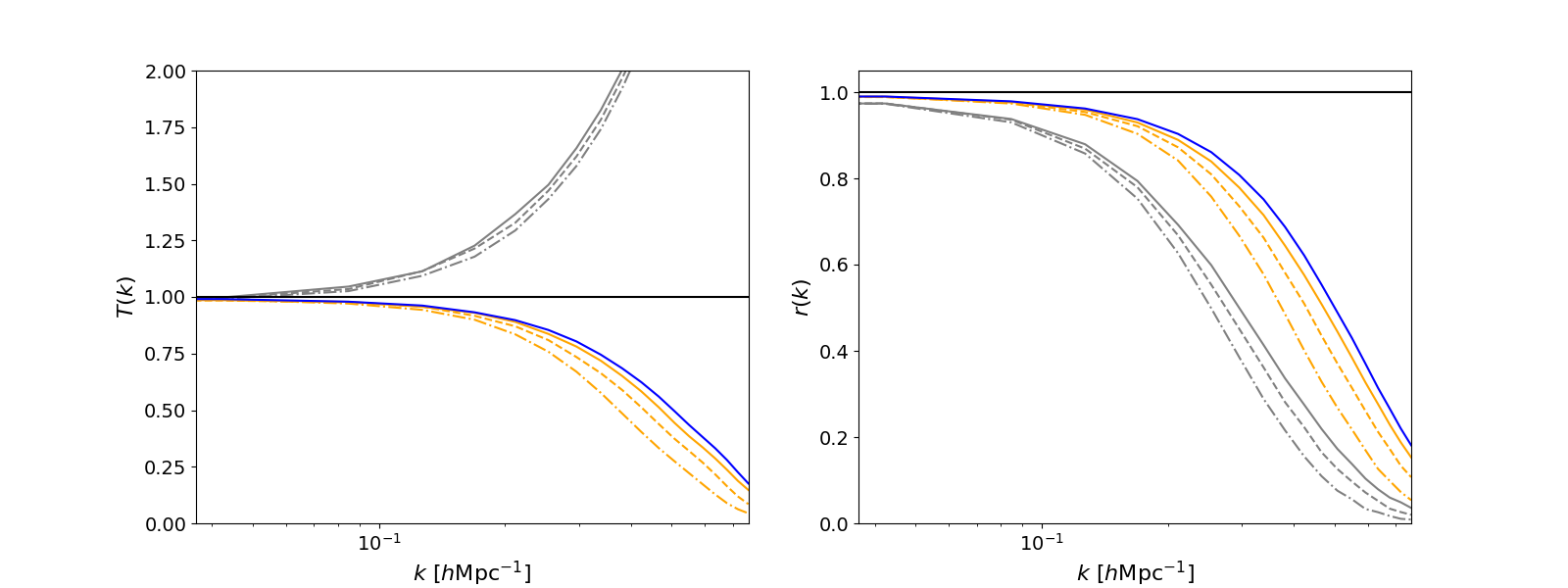}
        \caption{Mass-Weighted Halos}
    \end{subfigure}    
    \begin{subfigure}[b]{\textwidth}
        \centering
        \includegraphics[width=\textwidth]{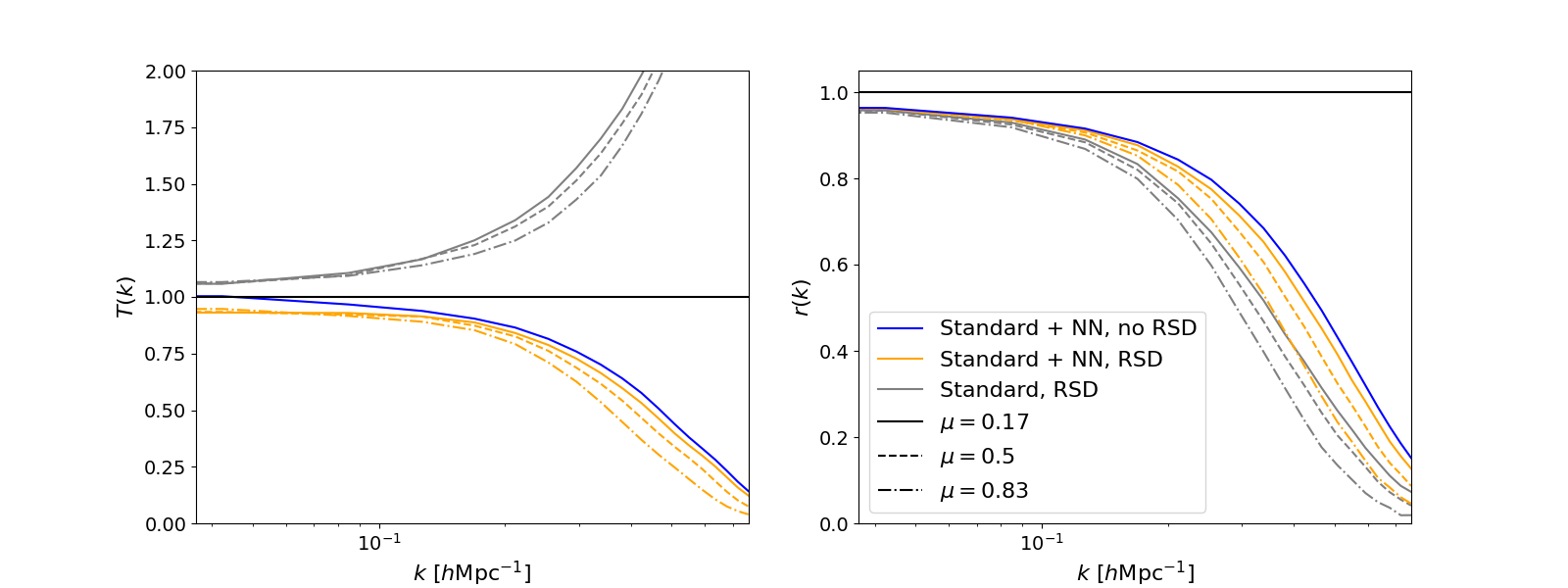}
        \caption{Number-Weighted Halos}
    \end{subfigure}
    \caption{\textbf{Redshift-space distortion (RSD) analysis for halo catalogs at $\mathbf{\bar n = 5\times 10^{-4}}$.} In the top panel, the network takes as input mass-weighted halos, while in the bottom panel, it takes as input number-weighted halos. In each sub-panel, linestyles correspond to different \(\mu\)-bins ranging from mostly transverse modes (\(\mu \approx 0.17\)) to nearly line-of-sight modes (\(\mu \approx 0.83\)) relative to the RSD direction. Orange lines show results from our CNN-corrected reconstruction, while gray lines depict the standard reconstruction, and the blue line indicates the reconstruction performance on configuration-space fields. Our method's improvement over standard reconstruction is consistent even at higher \(\mu\) where redshift-space effects dominate.}
    \label{fig:rsd_analysis}
\end{figure*}

Figure~\ref{fig:rsd_analysis} plots these results for several ranges of $\mu$, from nearly transverse modes centered on $\mu \approx 0.17$ to nearly line-of-sight modes centered on $\mu \approx 0.83$; all $\mu$ bins average over a $\mu$-range of $0.33$. Across all \(\mu\)-bins, our CNN-corrected reconstruction (orange) consistently achieves higher correlation and more accurate amplitude than standard reconstruction (gray) on intermediate to small scales \((k \geq 0.1\,h\,\mathrm{Mpc}^{-1})\), while converging at large scales. In fact, the CNN-corrected prediction in almost fully line-of-sight modes is still better than the standard reconstruction in almost fully transverse modes. As expected, redshift-space distortions degrade performance in modes aligned with the line of sight more strongly than those 
perpendicular to it. 

\subsubsection{Halo Mass Scatter}
\label{sec:halo-scatter}
We also explore how uncertainties in halo mass estimates affect the reconstruction. Specifically, we introduce a controlled level of noise in our mass-weighted samples by drawing random offsets \(\Delta M\) from a normal distribution and adding these to the halo masses, resulting in a root-mean-square (RMS) scatter spanning values from \(0.0\) up to \(1.3\). This procedure mimics imperfect mass determinations in observational data, which can arise from uncertainties in richness-based mass proxies, velocity dispersions, or photometric measures \citep{cai2011optimal, old2015galaxy, castro2016constraining}. We then evaluate our trained model on these mass-scattered mocks in redshift space. Importantly, we do not retrain our mass-weighted correction on the mass scattered mocks, but just observe the degradation in performance when faced with noisy halo masses.  

\begin{figure*}
    \centering
    \includegraphics[width=0.9\textwidth]{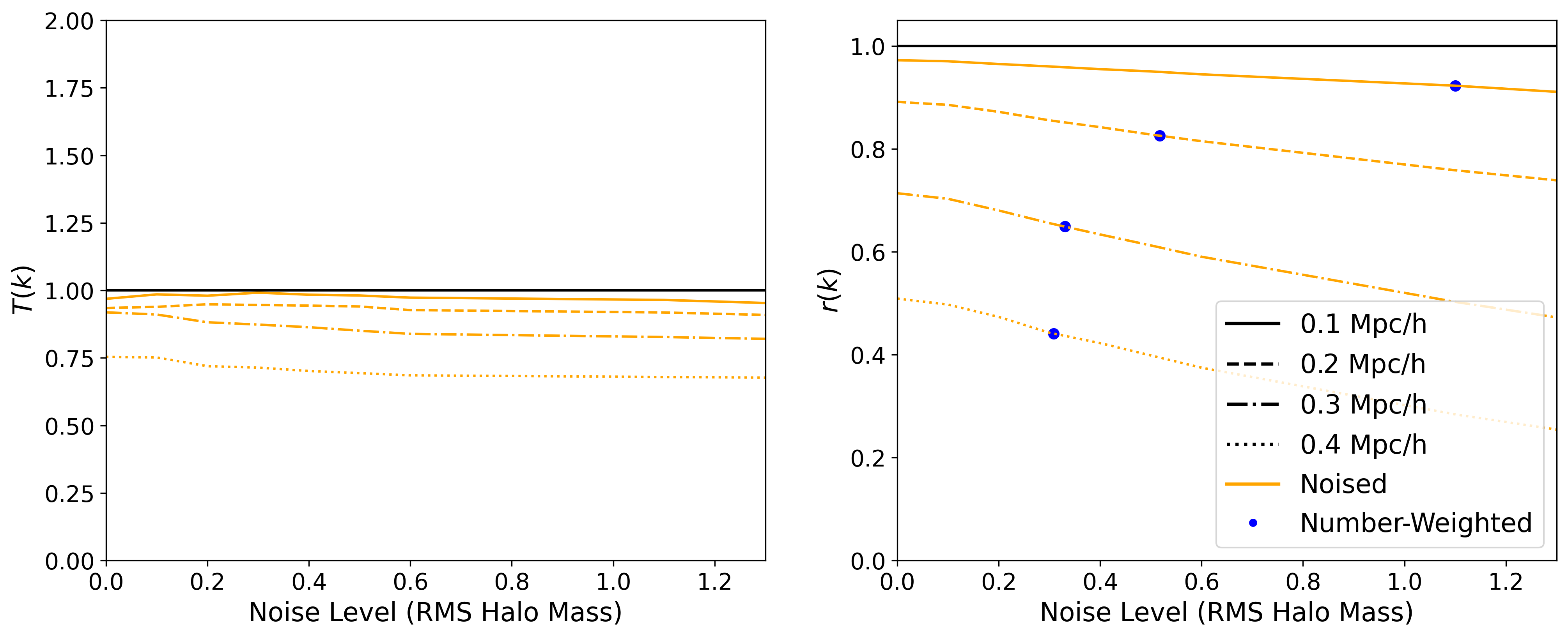}
    \caption{\textbf{Impact of halo mass scatter on reconstruction performance in redshift space at $\mathbf{\bar n = 5 \times 10^{-4}}$.} The horizontal axis indicates the RMS scatter added to each halo’s mass (from 0.0 up to 1.1), and the vertical axis plots the resulting transfer function ($T(k)$, left) or cross-correlation ($r(k)$, right) at characteristic scales \(k = 0.1,\,0.2,\,0.3,\,0.4\,h^{-1}\,\mathrm{Mpc}\). Orange lines (“Noised”) show how the reconstruction degrades when masses are randomly perturbed, whereas blue points (“Number-Weighted”) depict uniform-weighted halos. Notably, \cite{sullivan2023learning} report that using observables such as color and galaxy positions one can achieve mass uncertainties around \(\text{RMS}\!\approx\!0.24\) for future surveys, placing them in the regime where a reliable mass proxy still confers appreciable gains.}
    \label{fig:halo-scatter}
\end{figure*}

Figure~\ref{fig:halo-scatter} summarizes these results. Each curve depicts the reconstructed cross-correlation \(r(k)\) at a characteristic wavenumber (\(k = 0.1,\,0.2,\,0.3,\,0.4\,h^{-1}\,\mathrm{Mpc}\)) as a function of the injected RMS mass scatter. The orange lines (“Noised”) show the performance when we artificially degrade the mass estimates before applying our CNN-corrected reconstruction, while the blue points (“Number-Weighted”) indicate the performance of the uniform number-weighted model. As expected, increasing the scatter gradually erodes the benefits of mass weighting, reducing the cross-correlation on smaller scales. 

Nevertheless, for moderate scatter levels (RMS \(\leq 0.3\)), mass weighting continues to outperform uniform weighting—indicating that the reconstruction still benefits from approximate mass information, even when it is imperfect. In contrast, once the scatter grows beyond that point, performance begins to degrade below uniform-weighting at small scales. Notably, \cite{sullivan2023learning} report that future galaxy surveys can achieve halo mass uncertainties at the level of \(\sigma_{\rm RMS} \approx 0.24\). This suggests that real-world applications could remain firmly in the regime where mass-weighted reconstructions confer a clear advantage over uniform weighting.

\subsection{Galaxy Catalogs}
\label{sec:galaxy_results}

\subsubsection{Configuration Space Results}

\begin{figure*}
    \centering
    \begin{subfigure}[t]{\textwidth}
        \centering
        \includegraphics[width=\textwidth]{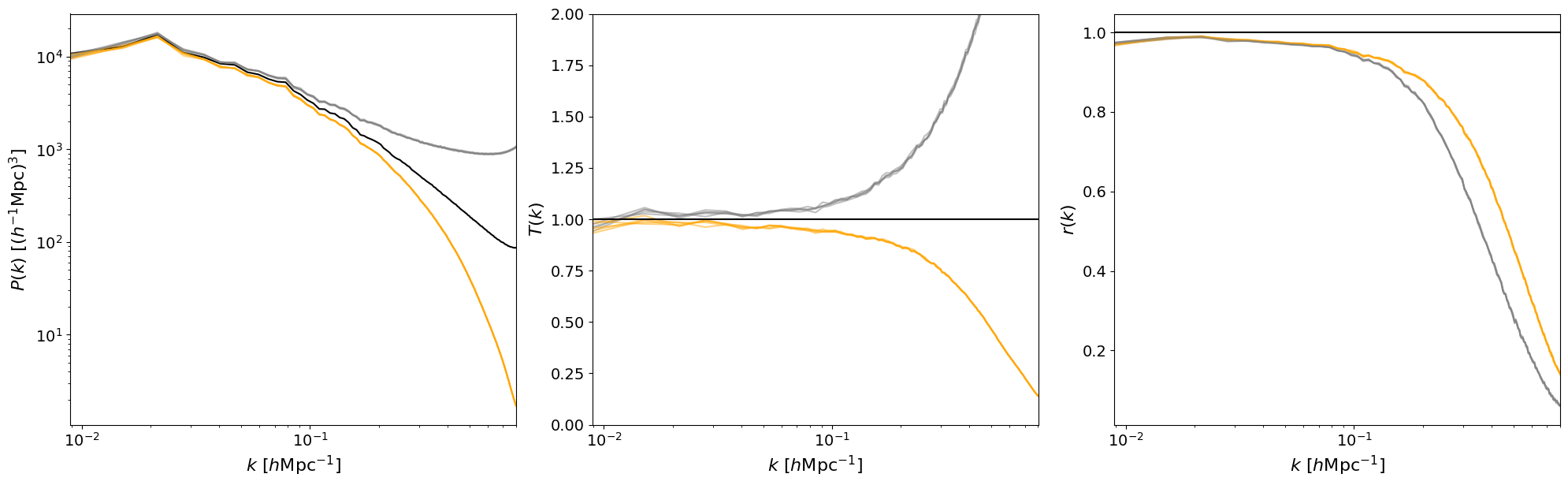}
        \vspace{-0.5cm}
        \caption{$\bar n = 1 \times 10^{-3}$}
        \vspace{0.5cm} 
    \end{subfigure}
    \begin{subfigure}[t]{\textwidth}
        \centering
        \includegraphics[width=\textwidth]{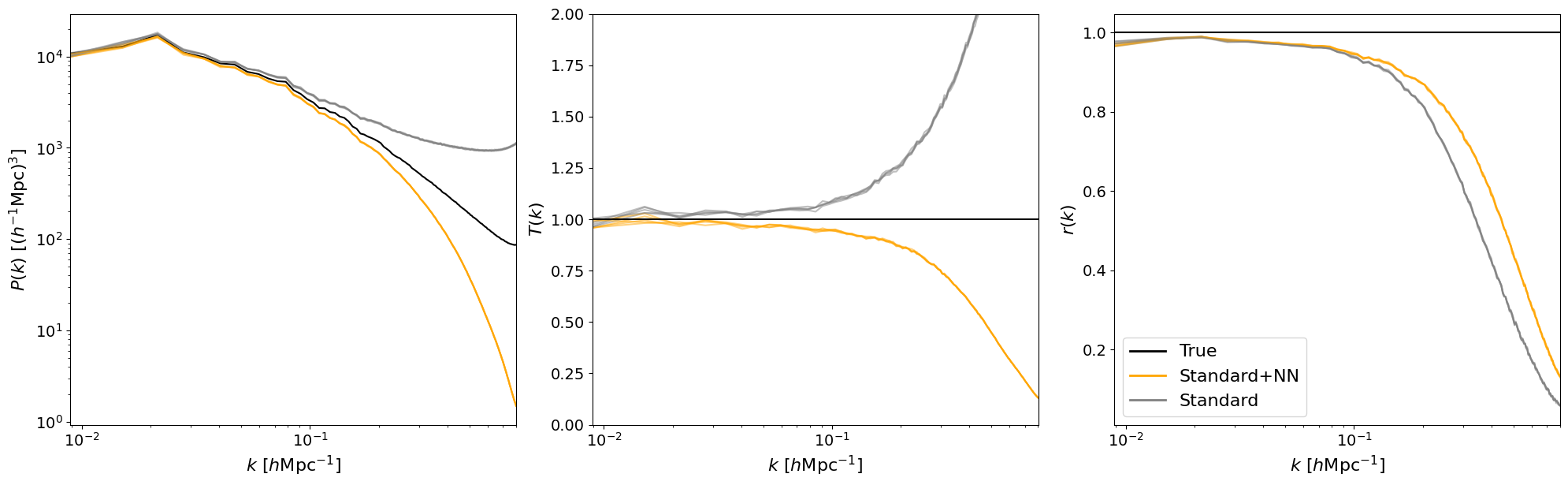}
        \vspace{-0.5cm}
        \caption{$\bar n = 5 \times 10^{-4}$}
        \vspace{0.5cm}
    \end{subfigure}
    \caption{\textbf{Reconstruction of the linear dark matter field from galaxies at \(\mathbf{z=0.5}\) in configuration space for two number densities.} For each primordial density field, we show reconstructions from galaxy fields generated using five different HOD parameter draws within the \cite{hahn2023simbig} HOD model. Standard reconstruction is shown in gray, while our CNN-corrected approach is shown in orange. Notably, the increased stochasticity arising from the HOD—such as the assignment of satellites and variability in halo-occupation thresholds—does not significantly degrade performance compared to uniformly weighted halos.}
    \label{fig:galaxy_config}
\end{figure*}

Having demonstrated our model's performance on dark matter halos, we now extend our pipeline to simulated galaxy fields. We retrain the same models as above on the galaxy fields detailed in \S\ref{sec:dataset} and present the results below.

Figure~\ref{fig:galaxy_config} compares our CNN-corrected reconstruction (orange) to standard reconstruction (gray) for galaxy catalogs at the two number densities used previously, \(\bar{n} = 1\times10^{-3}\) (top, DESI ELGs) and \(\bar{n} = 5\times10^{-4}\) (bottom, DESI LRGs). We show reconstructions of the same initial conditions from five different galaxy fields generated using the nine HOD parameters detail in Table \ref{tab:hod_params}. Similar to the halo analyses, the CNN corrections accurately recover both the amplitude and phase of the true initial field on intermediate to small scales, consistently outperforming the standard reconstruction. Additionally, we note that the model remains robust to varying HOD parameter draws, corresponding to varying satellite fractions and central-galaxy thresholds. Finally, the increased stochasticity arising from the HOD—such as the assignment of satellites and variability in halo-occupation thresholds—does not significantly degrade performance compared to uniformly weighted halos. Ultimately, these results underscore the model's effectiveness in inferring non-linear, small-scale features from the galaxy distribution and in mitigating non-linear galaxy bias shot noise. 

\subsubsection{Redshift-Space Distortions}

\begin{figure*}
    \centering
    \includegraphics[width=\textwidth]{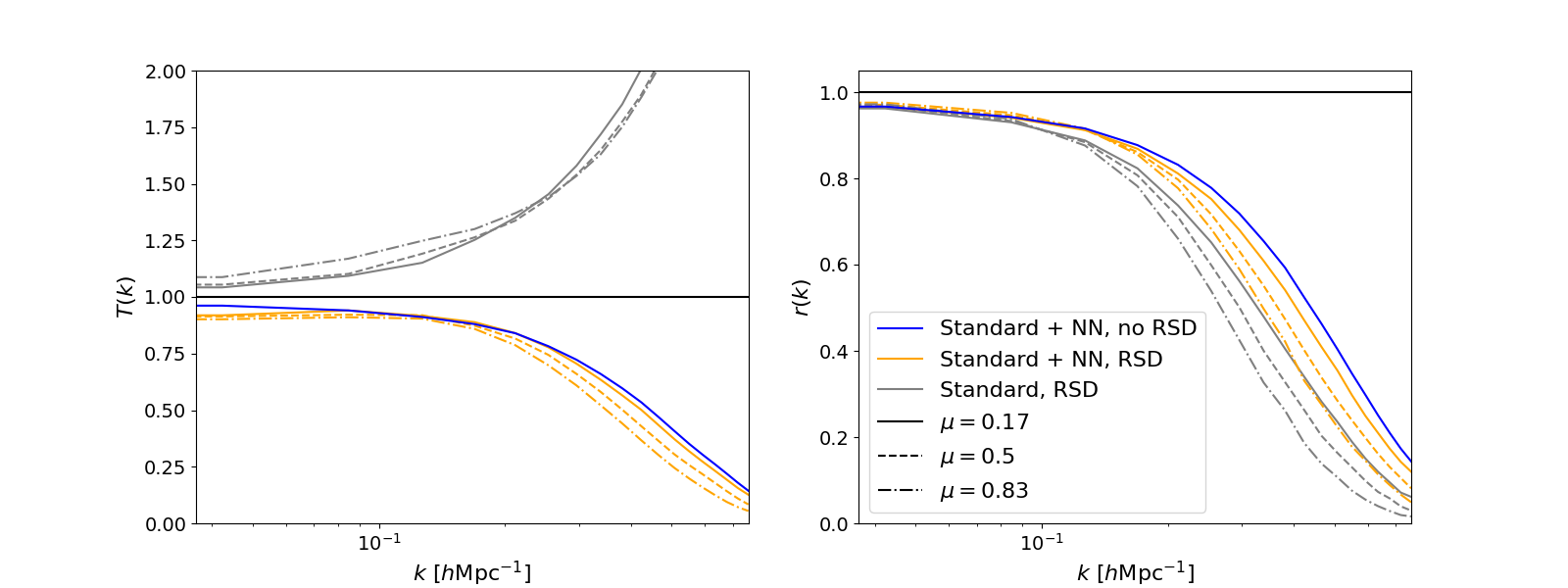}
    \caption{\textbf{Redshift-space distortion analysis for galaxy catalogs with $\mathbf{\bar n = 5\times 10^{-4}}$.} We show the anisotropic transfer function ($T(k)$, left) and cross-correlation coefficient (\(r(k,\mu)\), right) with the true primordial dark matter field in bins of wavenumber \(k\) and angle \(\mu\) relative to the line of sight for the number density \(\bar{n} = 5\times10^{-4}\). As with the halo-based RSD analyses, standard reconstruction is shown in gray, while our CNN-corrected approach is shown in orange, and the configuration-space CNN-correction is shown in blue for reference. Across all \(\mu\)-bins, the CNN-correction recovers the primordial density more accurately than the baseline.}
    \label{fig:galaxy-rsd}
\end{figure*}

As with halos, we bin Fourier modes by their magnitude \(k\) and angle \(\mu\) relative to the line of sight, and compute the anisotropic versions of the transfer function \(T(k)\) and the cross-correlation coefficient \(r(k)\) with the true primordial density field within each \((k,\mu)\)-bin. Once again, we use $\mu$-bins of width $\Delta \mu = 0.33$ centered on $\mu \in \{0.17, 0.5, 0.83\}$ as well as models trained specifically on the redshift-space fields instead of those trained in configuration space. The results are presented in Figure~\ref{fig:galaxy-rsd}. As with halos, across all $\mu$-bins, our CNN-corrected reconstruction (orange) achieves higher correlation than standard reconstruction (gray), with redshift-space distortions degrading performance in modes aligned with the line of sight more strongly than those perpendicular to it. 

\subsection{Scaling to Arbitrarily Large Volumes}
\label{sec:scaling_analysis}

So far in this analysis, we have presented results from our CNN correction, which we have trained on \(195\,h^{-1}\,\mathrm{Mpc}\) sub-volumes within a \(1\,h^{-1}\,\mathrm{Gpc}\) box, on held-out simulations that are themselves also in \(1\,h^{-1}\,\mathrm{Gpc}\) boxes. However, because the CNN correction processes each sub-volume independently, the CNN can be applied to arbitrarily large boxes without any additional retraining.  

To demonstrate this use case, we use a model trained on sub-volumes within a \(1(h^{-1}\,\mathrm{Gpc})^3\) box to reconstruct primordial density fields for simulations with box sizes of \(3(h^{-1}(\mathrm{Gpc})^3\) corresponding to $768^3$ voxels. The results presented in this section are from models trained on subgrids from the standard-resolution \(1(h^{-1}\textrm{Gpc})^3\) \textsc{Quiijote} fiducial simulations, as the \(3(h^{-1}(\mathrm{Gpc})^3\) suite is itself run at standard-resolution (rather than the high-resolution simulations used previously). They are then applied directly to the \(3(h^{-1}\,\mathrm{Gpc})^3\) simulations. All training and evaluation in this section is done directly in redshift-space, where redshift-space positions and the line of sight are consistent with the previous analyses presented above.

\begin{figure*}[t]
    \centering
    %
    \begin{subfigure}[b]{0.3\textwidth}
        \centering
        \includegraphics[width=\textwidth]{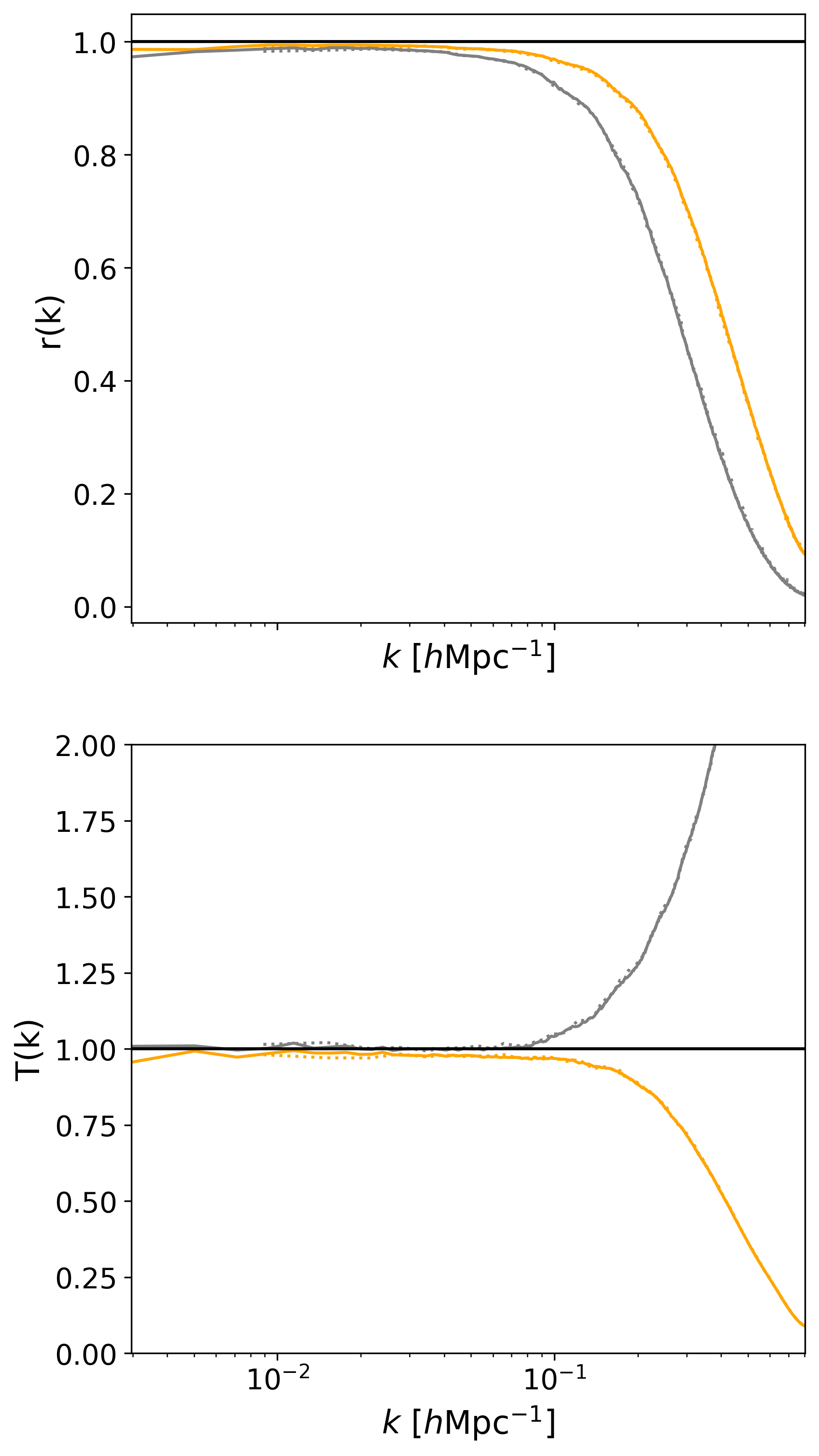}
        \caption{Halos (Mass-Weighted)}
    \end{subfigure}%
    %
    \begin{subfigure}[b]{0.3\textwidth}
        \centering
        \includegraphics[width=\textwidth]{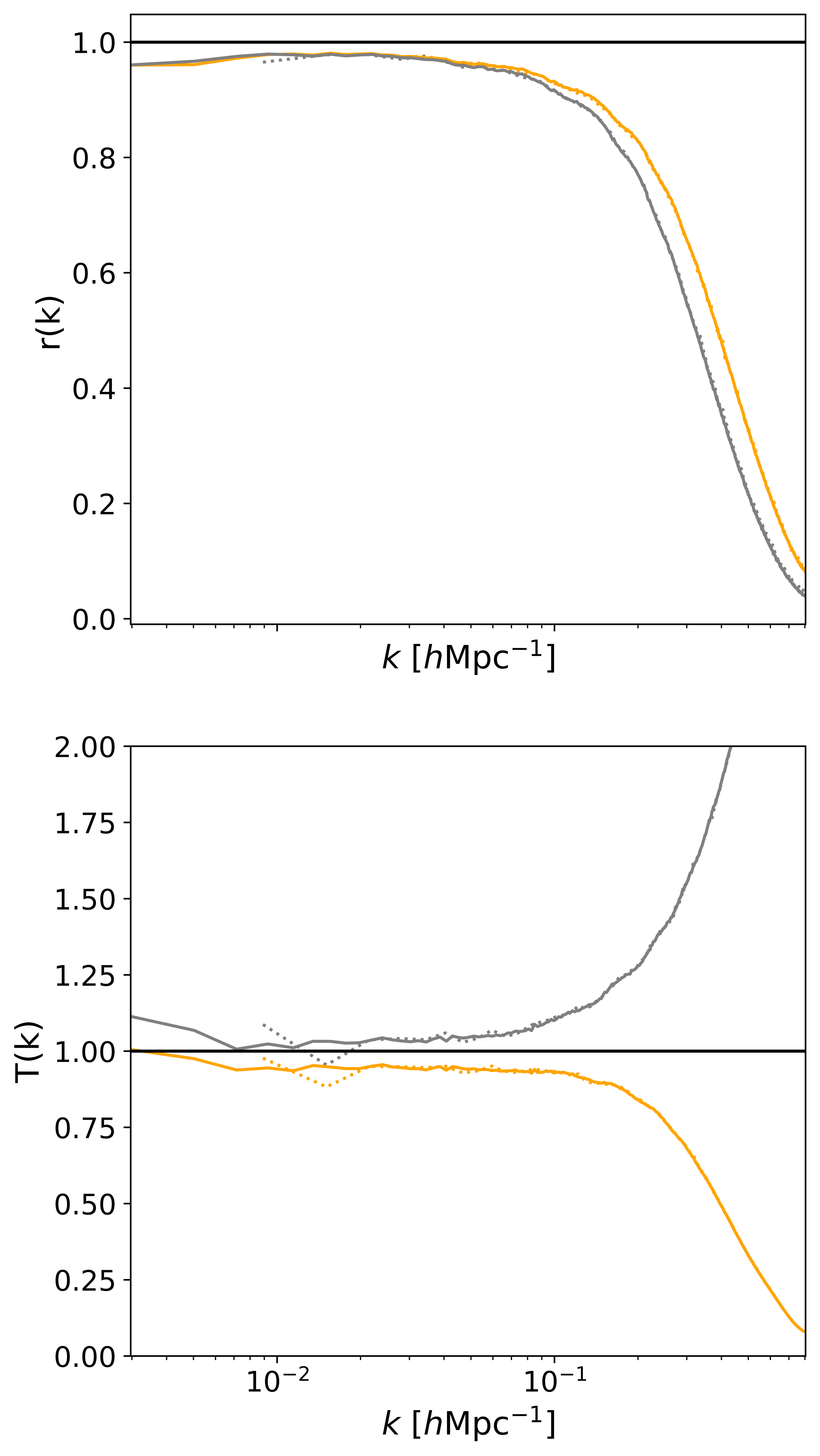}
        \caption{Halos (Number-Weighted)}
    \end{subfigure}%
    %
    \begin{subfigure}[b]{0.3\textwidth}
        \centering
        \includegraphics[width=\textwidth]{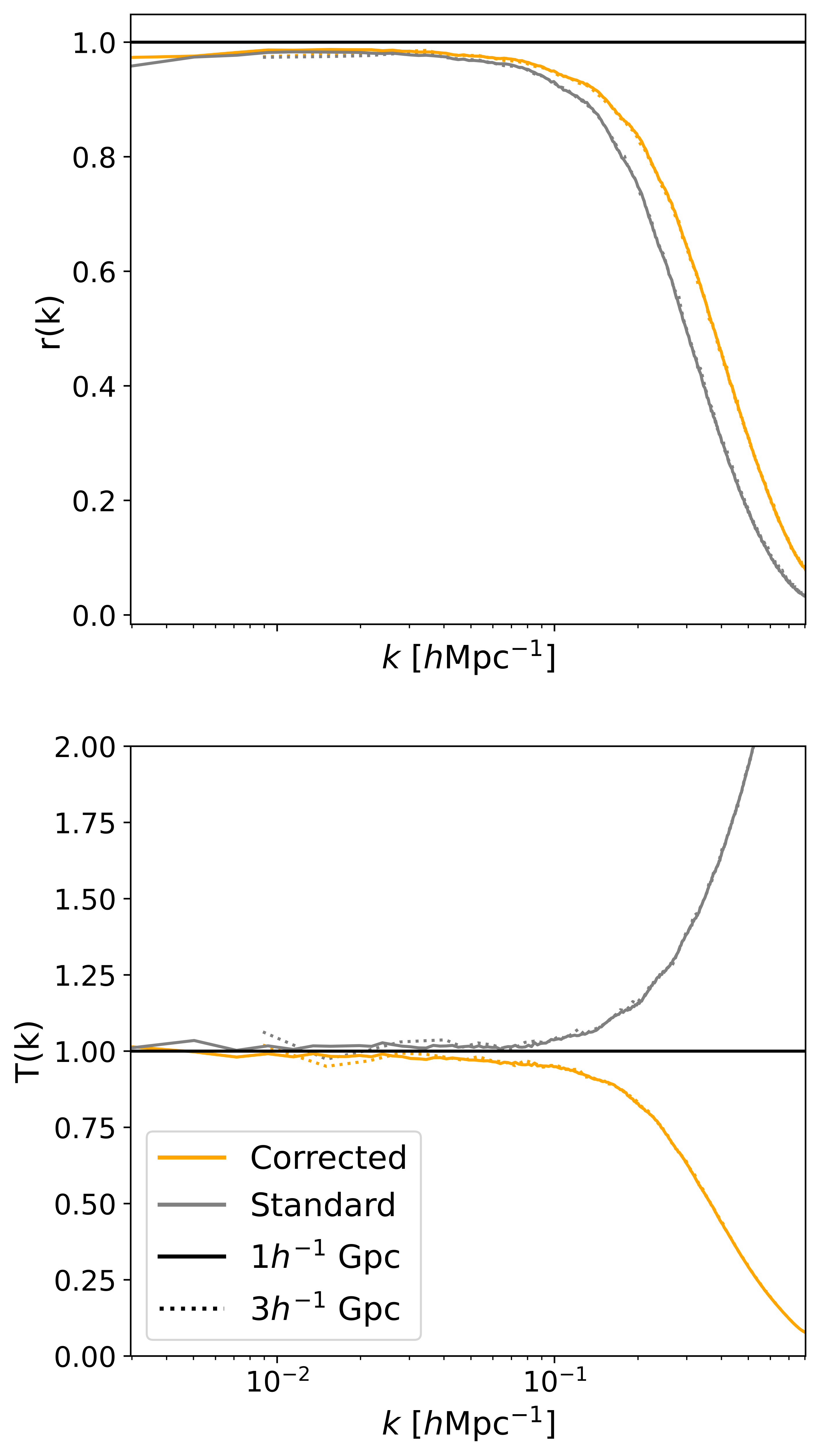}
        \caption{Galaxies}
    \end{subfigure}
    \caption{\textbf{Generalization to larger simulation volumes without retraining.} While our CNN is trained on \(\sim 200\,h^{-1}\,\mathrm{Mpc}\) sub-volumes within a \(1\,h^{-1}\,\mathrm{Gpc}\) box, we can directly apply it---without retraining---to significantly larger volumes. Here, we apply the CNN reconstruction to $3h^{-1}\mathrm{Gpc}$ boxes ($768^3$ voxels) for mass-weighted halos (left), number-weighted halos (middle), and galaxies (right), all at a number density of $\bar n = 5 \times 10^{-4}$ in redshift-space. We present the transfer function ($T(k)$, bottom) and the cross-correlation coefficient ($r(k)$, top) for both the $1h^{-1}\mathrm{Gpc}$ (solid) and $3h^{-1}\mathrm{Gpc}$ (dotted) reconstructions. Ultimately, there is no discernible loss in performance when generalizing to larger survey volumes.}
    \label{fig:scaling}
\end{figure*}

We present the results of these tests, for both dark matter halos and galaxies, in Figure \ref{fig:scaling}. We observe that the cross-correlation remains consistent at larger volumes, indicating that our sub-grid corrections remain robust. In practical terms, this capability is especially valuable for near-future experiments like DESI, which will extremely large survey volumes. Rather than needing bespoke large simulations and retraining for each new survey geometry, we can rely on a single sub-volume–trained network, thus markedly reducing computational overhead while preserving high fidelity in the recovered primordial density fields.

\subsection{Robustness to Model Misspecification} 
\label{sec:model-mis}
Simulation-based inference (SBI) relies on mocks that accurately encapsulate the forward model of the observables. In practice, the true observables may deviate from the assumed model prescription—an effect that can degrade predictions if the simulation training set is not well matched to reality. To gauge the impact of such model misspecification, we examine two scenarios. First, we assess performance on mocks generated with a different halo occupation distribution (HOD) than that used for training. Second, we train on halo fields but apply the results to galaxy catalogs. These tests demonstrate how robust our CNN-corrected approach remains when the forward modeling in the training set does not exactly match the test data.

\subsubsection{Different Halo Occupation Distribution}
\label{sec:different-hod}
\begin{figure*}
    \centering 
    \includegraphics[width=0.9\textwidth]{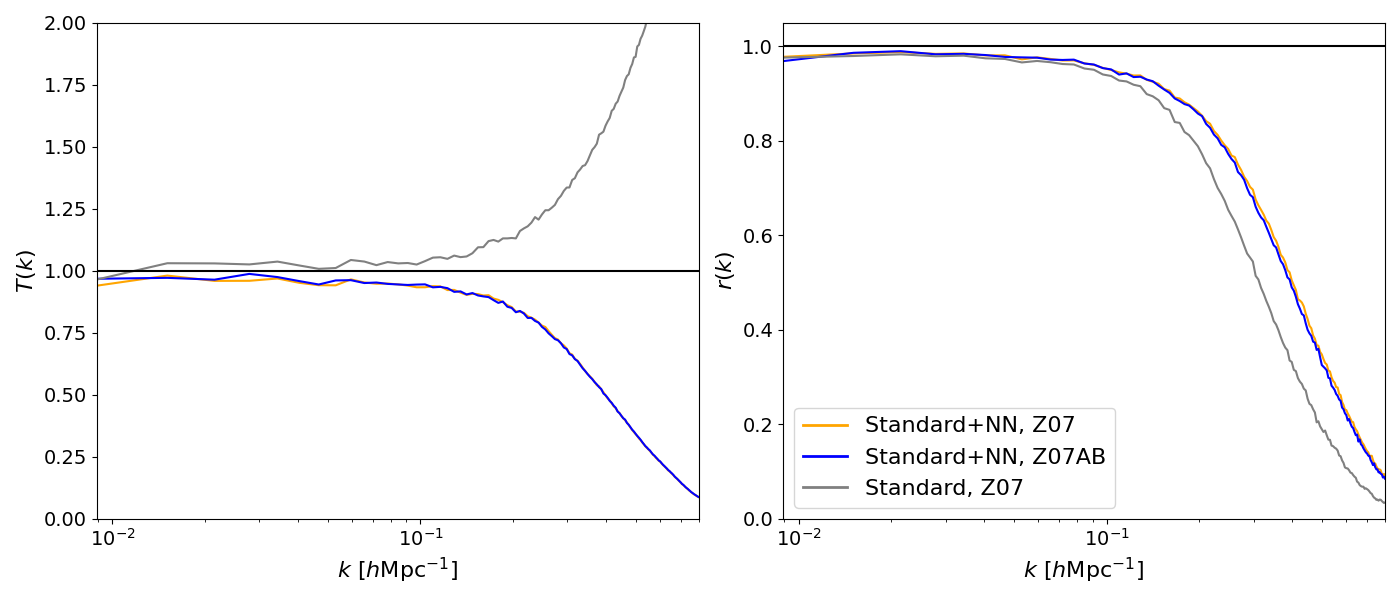} 
    \caption{\textbf{Robustness to HOD Misspecification.} We train on the Z07AB HOD, which includes velocity, assembly, and concentration biases, and evaluate on mocks generated with the standard Z07 HOD \cite{zheng2007galaxy} (blue). For comparison, we also show a network trained and tested fully on the Z07 HOD (orange). The panels display the transfer function ($T(k)$, left) and cross-correlation coefficient ($r(k)$, right) for a redshift-space galaxy sample at $\bar{n} = 5 \times 10^{-4}$. Despite the HOD mismatch, the Z07AB--Z07 combination (blue) largely agrees with the fully consistent Z07--Z07 setup (orange). We conclude that modest deviations in the galaxy--halo connection do not significantly degrade CNN performance.} 
    \label{fig:hod_mis} 
\end{figure*}

We first test whether our CNN’s local, small-scale corrections degrade under modest changes in galaxy occupation physics. The Z07AB HOD (used for training in earlier sections) extends the standard Z07 \cite{zheng2007galaxy} model by incorporating assembly bias, velocity bias, and concentration bias. Here, without retraining, we evaluate that same CNN on mocks using only the base Z07 HOD. Figure~\ref{fig:hod_mis} compares these reconstructions (blue) to a network both trained and tested on the simpler Z07 model (orange). The results indicate that a moderate HOD mismatch between training and testing data has negligible impact on performance; the Z07AB--Z07 CNN reconstructions remain nearly as accurate as their fully consistent counterparts.

\subsubsection{Halos to Galaxies}

\begin{figure}
    \centering 
    \includegraphics[width=0.9\textwidth]{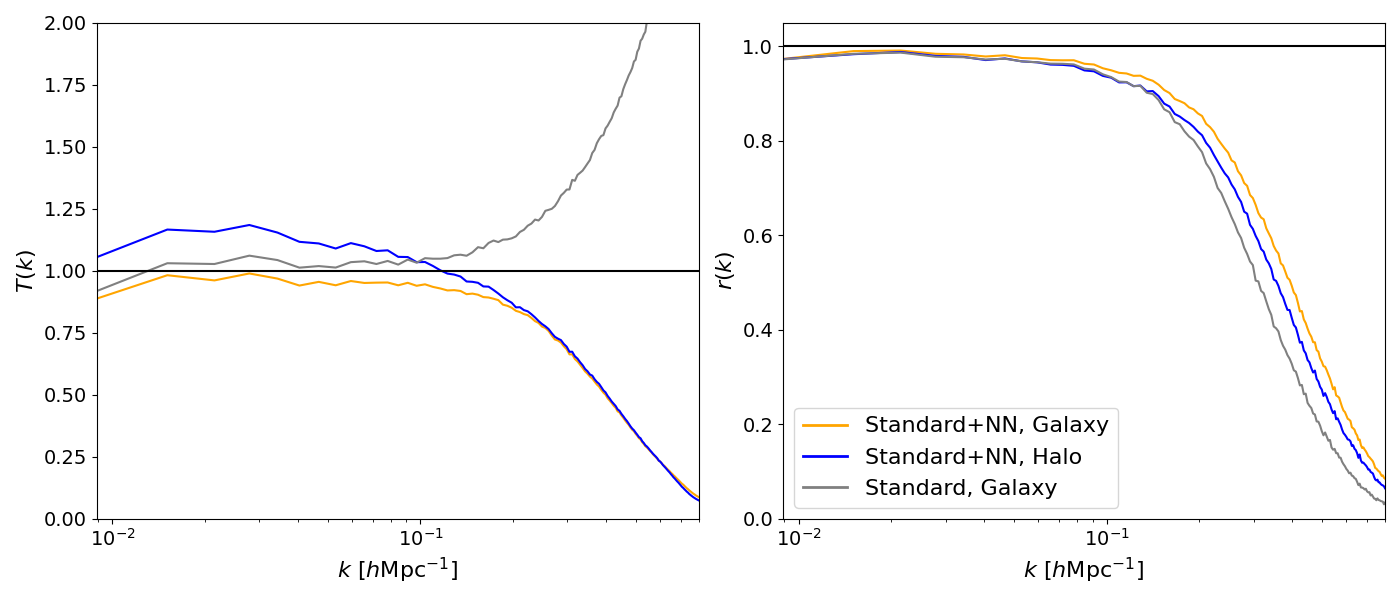} 
    \caption{\textbf{Training on Halos, Inferring on Galaxies.} We train the CNN corrections on halo fields (in redshift space) but evaluate on redshift-space galaxy mocks with the same number density, $\bar{n} = 5 \times 10^{-4}$. We present the transfer function ($T(k)$, left) and the cross-correlation coefficient ($r(k)$, right) with the true initial dark matter field. As shown by the blue curves, although this “halo-trained” model is not as accurate as a galaxy-trained one (orange), it still outperforms standard reconstruction (gray). This partial transferability indicates that the network learns useful small-scale corrections even without detailed galaxy--halo relationships. However, a direct galaxy-based training remains optimal when possible.} 
    \label{fig:halo2gal_fig} 
\end{figure}

Next, we examine the effects of training a network on halo fields but then applying it directly to galaxy catalogs at the same number density with no retraining. Figure~\ref{fig:halo2gal_fig} compares these “halo-to-galaxy” reconstructions (blue) against a CNN trained directly on galaxy mocks (orange) and evaluated on galaxy mocks. Although the halo-trained network loses some fidelity on small scales—where satellite fractions and HOD stochasticity matter—its reconstructions are still better than standard reconstruction (gray) in terms of cross-correlation. This suggests that even a simplified forward model can capture enough local physics to improve upon traditional methods. However, training directly on galaxy mocks is recommended for fully accounting for HOD-related scatter, especially given that the network appears to be robust to moderate HOD model misspecification as demonstrated in \S\ref{sec:different-hod}.

\section{Fisher Analysis of the BAO Peak}
\label{sec:bao-info}

A central objective in BAO analyses is to quantify how well the acoustic scale 
distance can be measured from a given density field.  In the usual Fisher-matrix formalism, one considers how the power spectrum \(P(\mathbf{k})\) shifts under small perturbations to the distance to the sound horizon at the drag epoch, \(s_0\). Assuming that the band powers of the power spectrum follow a Gaussian likelihood \citep{seo2007improved}, we write the Fisher matrix element for \(\ln s_0\) as
\begin{equation}
F_{\ln s_0}
\;=\; V_{\mathrm{sur}}
 \int_{k_\mathrm{min}}^{k_\mathrm{max}} 
    \!\!
    \left(\frac{\partial P_{\mathrm{pred,pred}}(k)}{\partial \ln s_0}\right)^2
    \; \frac{1}{2[P_{\mathrm{pred,pred}}(k)]^2} 
    \,\frac{4\pi k^2 \, dk}{(2\pi)^3},
\label{eq:flns0-generic}
\end{equation}
where \(V_{\mathrm{sur}} = 1\,(h^{-1}\mathrm{Gpc})^3\) in our setup. In a general model where we are 
reconstructing initial 
conditions encoded in 
$P_{\mathrm{truth,truth}}(\mathbf{k})$
we can relate 
cross-correlations and 
auto-correlations by 
introducing a noise term,
\begin{align}
 P_{\mathrm{pred,pred}}(\mathbf{k})=\left(\frac{P_{\mathrm{pred,truth}}(\mathbf{k})}{P_{\mathrm{truth,truth}}(\mathbf{k})}\right)^2[P_{\mathrm{truth,truth}}(\mathbf{k})+N(\mathbf{k})],  
\end{align}
where $N(\mathbf{k})$ is the noise power spectrum of reconstruction. Inserting 
this into the cross-correlation coefficient definition of Eq. \ref{rcc} one obtains 
$r^2(\mathbf{k})=(1+N(\mathbf{k})/P_{\mathrm{truth,truth}}(\mathbf{k}))^{-1}$. We can now write the Fisher matrix as
\begin{equation}
    F_{\ln s_0} =  \frac{V_{\rm sur}}{(2\pi)^2}\int_{k_{\mathrm{min}}}^{k_{\mathrm{max}}} 
    \left(\frac{ d\ln P_{\rm{truth,truth}}}{d\ln s_0}\right)^2 r(k)^4
    k^2dk
\label{eq:F_ln_s0-app}
\end{equation}
We see from this expression that the error on the distance scale depends only on the cross-correlation coefficient $r(k)$, in addition to the response of the linear power spectrum to the BAO distance scale.
In this derivation we have 
implicitly assumed that 
there is no residual nonlinear BAO wiggle smoothing left in the 
CNN reconstruction, in 
contrast to the 
standard BAO reconstruction which does not completely
eliminate nonlinear BAO 
smoothing. 

Following \cite{seo2007improved}, we note that the BAO can be approximated by \(\ j_0(ks_0)=\sin(ks_0)/(ks_0)\), and broadening of the peak from Silk damping or non-linear damping appears as exponential suppression in Fourier space. Therefore, the ``baryonic wiggle'' component can be written as proportional to the original \( j_0(ks_0)\) factor multiplied by damping envelopes for these types of damping. This is equivalent to the BAO Fisher analysis presented in \cite{modi2018cosmological}, which gives
\begin{equation}
    F_{\ln s_0} = V_{\rm sur} A_0^2 \int_{k_{\mathrm{min}}}^{k_{\mathrm{max}}} e^{-2(k \Sigma_\mathrm{s})^{1.4}}r(k)^4 k^2 \left( \frac{P_{\mathrm{truth,truth}}(0.2)}{P_{\mathrm{truth,truth}}(k)}\right)^2 dk = \left( \frac{s_0}{\sigma_{s_0}} \right)^2.
\end{equation}
Here, we set \(\Sigma_\mathrm{s} \simeq 7.76\,h^{-1}\,\mathrm{Mpc}\) for the effective Silk-damping scale and $s_0=150{\rm Mpc}/2\pi$ for the BAO scale. Additionally, we include the effects of non-linear damping for standard reconstruction by adding an additional damping envelope of $e^{-k^2\Sigma_{\mathrm{NL}}^2}$, where we set \(\Sigma_{\mathrm{NL}}=5.6 h^{-1}\textrm{Mpc}\) \cite{seo2007improved} 
as the average nonlinear damping remaining for
standard reconstruction 
in redshift space.

\begin{figure}[t]
    \centering
    \begin{subfigure}[t]{0.32\linewidth}
      \centering
      \includegraphics[width=\linewidth]{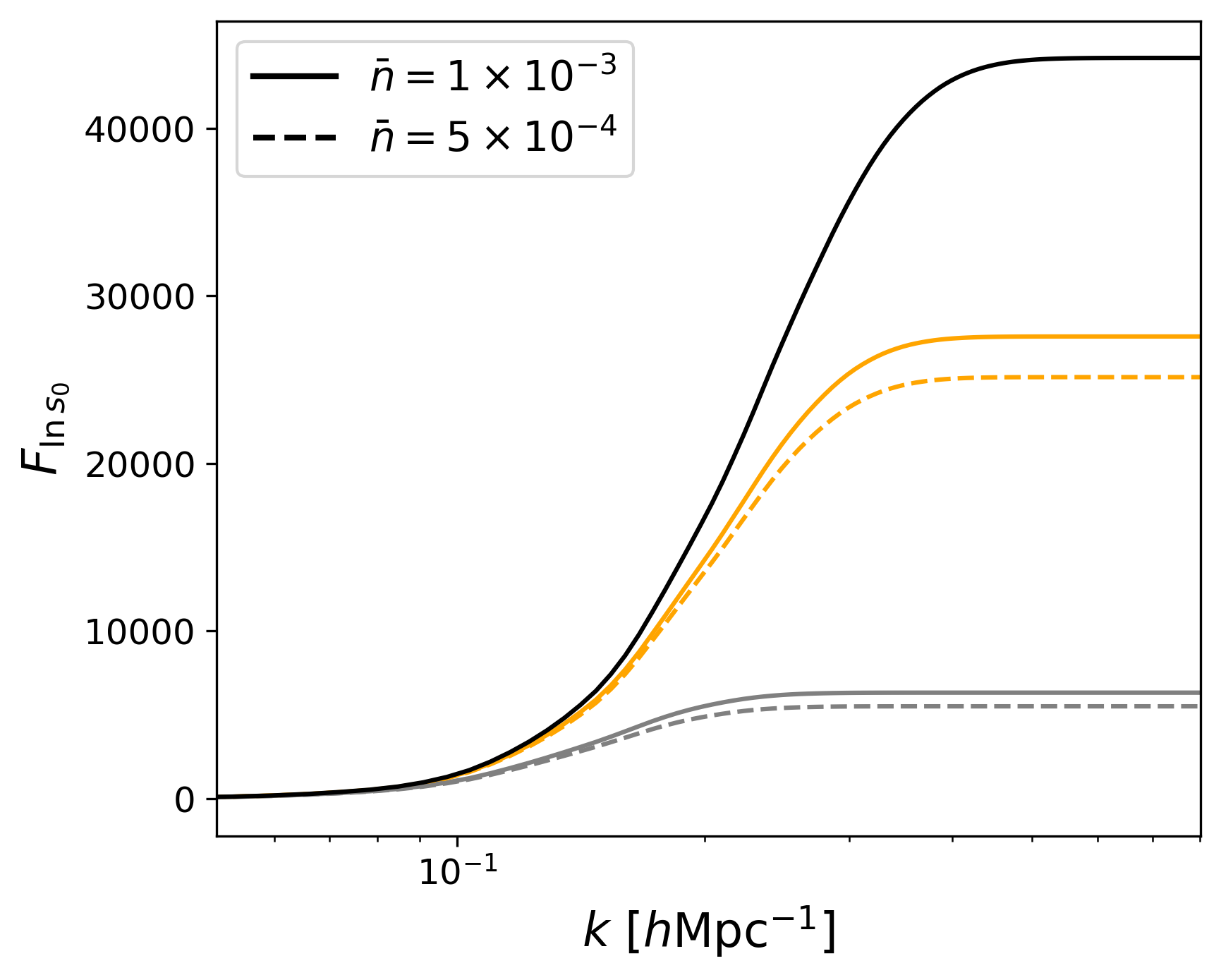}
      \caption{Halos (Mass-Weighted)}
    \end{subfigure}\hfill
    \begin{subfigure}[t]{0.32\linewidth}
      \centering
      \includegraphics[width=\linewidth]{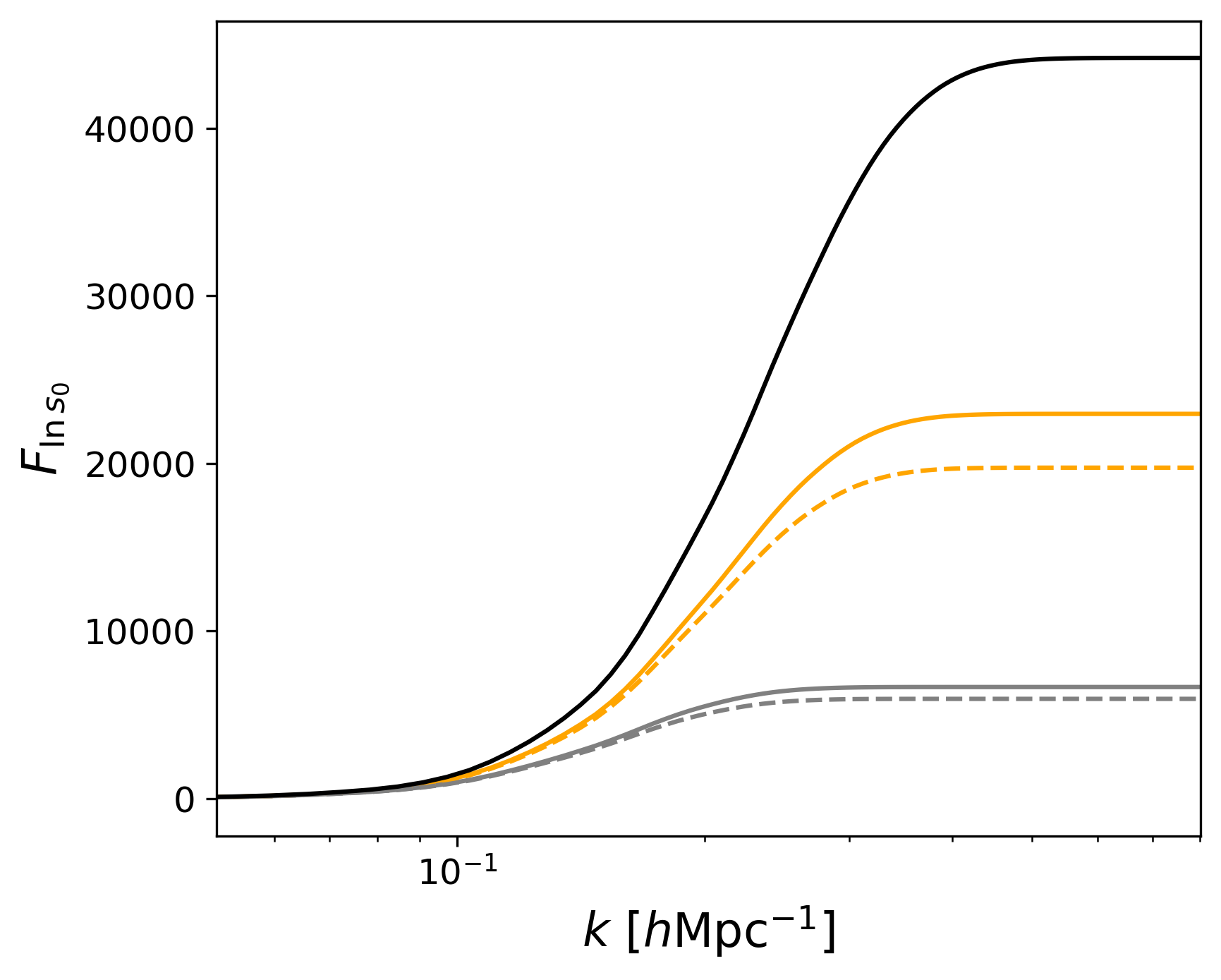}
      \caption{Halos (Number-Weighted)}
    \end{subfigure}\hfill
    \begin{subfigure}[t]{0.32\linewidth}
      \centering
      \includegraphics[width=\linewidth]{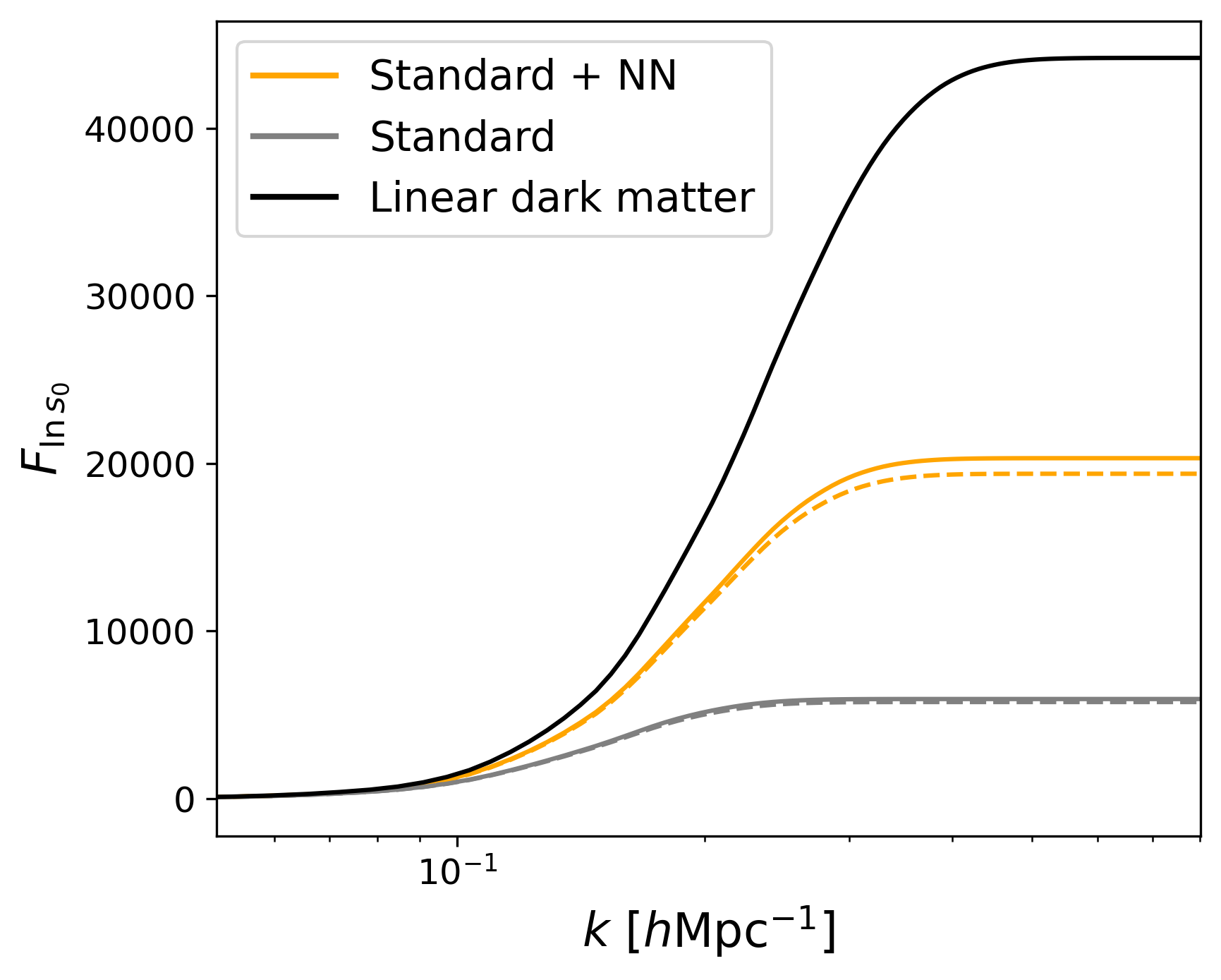}
      \caption{Galaxies}
    \end{subfigure}
    \caption{\textbf{Cumulative Fisher information on the BAO distance $\mathbf{\ln s_0}$.} We present $F_{\ln s_0} = (s_0/\sigma_{s_0})^2$ as a function of $k_{\mathrm{max}}$ for our CNN-corrected reconstructions (orange) and standard reconstruction (grey) at two input number densities, $\bar n = 5 \times 10^{-4}$ and $\bar n = 1 \times 10^{-3}$. We also include the Fisher information for the true linear field (black), which serves as a rough upper bound on the information limit present at a given $k$ value.}
    \label{fig:bao_info_figure}
\end{figure}

We evaluate the Fisher information as a function of $k_{\mathrm{max}}$ in Figure~\ref{fig:bao_info_figure}. We include the results for the CNN-corrected (orange) and standard reconstruction (grey) estimates of the primordial density field at the two number densities evaluated in the previous analysis, $\bar n = 5\times 10^{-4}$ (dotted) and $\bar n = 1 \times 10^{-3}$ (solid). We also include the Fisher information for the true linear field (black), which represents the natural upper bound on the Fisher information. Overall, we see that the CNN-corrected reconstruction materially outperforms standard reconstruction at all scales, for both halos and galaxies, and at both number densities. For example, when reconstructing the primordial density field from galaxies in redshift-space, the CNN correction achieves an error $\sigma_{s_0}/s_0 = 0.70\%$ at $\bar n = 1\times 10^{-3}$ and $0.71\%$ at $\bar n = 5 \times 10^{-4}$, while standard BAO reconstruction yields errors of $1.30\%$ and $1.32\%$ respectively; in other words, standard BAO reconstruction's error is $85\%$ larger than the CNN. This general trend holds true for both mass-weighted and number-weighted halos, although the discrepancy between errors at varying number densities is more significant, and the overall error is lower for number-weighted halos and lowest for mass-weighted halos, as expected. 

\section{Discussion}
\label{sec:discussion}
Our results demonstrate that coupling standard reconstruction with a CNN-based subgrid correction can yield significant improvements in recovering the primordial density field from biased tracers, surpassing what purely perturbative approaches achieve on nonlinear scales. 

For dark matter halos in both real-space and redshift-space analyses, our CNN-corrected reconstruction substantially improves upon standard techniques in recovering the linear density. In real space, the improved method recovers nearly the full large-scale amplitude and maintains a cross-correlation coefficient \(r(k) > 0.9\) out to \(k \sim 0.3\,h\,\mathrm{Mpc}^{-1}\) for mass-weighted halos, with the benefit most pronounced for higher-density catalogs. The advantage of mass weighting persists under moderate levels of mass scatter (up to \(\text{RMS}\sim 0.3\)) but drops below uniform weighting for large scatter. Even in redshift space, where line-of-sight modes are more strongly affected by non-linearities, our CNN-corrected approach consistently outperforms standard reconstruction across all bins in \(\mu\). Collectively, these results highlight the substantial gains in both amplitude recovery and phase accuracy that can be obtained by combining standard large-scale reconstruction with a trained neural network correction, particularly if reliable halo-mass estimates are available.

Applying the same CNN-based reconstruction framework to galaxy catalogs demonstrates that our method remains robust to the additional stochasticity of the halo occupation distribution. Compared to standard reconstruction, the CNN-corrected results exhibit improved cross-correlation on intermediate to small scales, even for different HOD parameter draws. Although the absence of direct mass information leads to somewhat lower overall correlations than in the mass-weighted halo case, the performance gain over the baseline remains substantial. This improvement persists in redshift space, where our approach again outperforms standard reconstruction across all angles to the line of sight. 

Moreover, tests of model misspecification reveal that the CNN is not overly sensitive to moderate changes in the assumed small-scale matter--tracer connection. Even if trained on an augmented HOD (including assembly, velocity, or concentration biases) but tested on a simpler HOD, or trained on halos but applied to galaxies, the reconstruction remains more accurate than the standard BAO reconstruction. This suggests that the learned subgrid corrections transfer reliably across modest variations in the forward model, alleviating concerns about overfitting to specific HOD prescriptions.

Finally, we quantify how these improvements translate into tighter BAO constraints by performing a Fisher-matrix analysis of the acoustic scale.  Across both halos and galaxies, and for different number densities, the CNN-corrected fields yield substantially more BAO information than standard reconstruction; for galaxies in redshift-space at the DESI LRG number density, for example, the CNN is able to achieve an error on the acoustic scale of $0.71\%$, while standard BAO reconstruction yields an error of $1.32\%$---the CNN reduces the error by $85\%$ relative to standard BAO reconstruction.

\begin{table}
\centering
\caption{\textbf{Approximate GPU memory usage requirements for one sample}. Requirements are estimated based on approximate activation size and model size for training a full-grid CNN or a typical full-grid U-Net on different input volumes with batch size 1. In our study, we instead train on subgrids of size $50^3$ voxels (\(\sim\)195$\,h^{-1}\,\mathrm{Mpc}$) to keep GPU usage manageable; we highlight the memory requirements of our method, which remain constant across any physical inference volume, in bold. For reference, a NVIDIA H100 typically has $80\,\mathrm{GB}$ of on-device RAM.}
\label{tab:memory_estimates}
\begin{tabular}{lcccc}
\toprule
\textbf{Voxels} & \textbf{Physical Size} & \textbf{CNN} & \textbf{U-Net} \\
\midrule
\(50^3\)   & \(195\,(h^{-1}\,\mathrm{Mpc})^3\)   & \(\mathbf{620.1\,\mathrm{\textbf{MB}}}\)    & \(64.2\mathrm{MB}\)   \\
\(256^3\)  & \(1\,(h^{-1}\,\mathrm{Gpc})^3\)     & \(73.4\,\mathrm{GB}\)   & \(8.6\,\mathrm{GB}\)   \\
\(768^3\)  & \(3\,(h^{-1}\,\mathrm{Gpc})^3\)     & \(945.2\,\mathrm{GB}\)  & \(232.3\,\mathrm{GB}\)  \\
\(2048^3\) & \(8\,(h^{-1}\,\mathrm{Gpc})^3\)     & \(43.4\,\mathrm{TB}\)   & \(4.6\,\mathrm{TB}\)   \\
\bottomrule
\end{tabular}
\end{table}

Compared to standard, full-box deep learning driven methods, one significant advantage of subgrid training is that it significantly mitigates GPU memory requirements: as indicated by Table~\ref{tab:memory_estimates}, end-to-end training on gigaparsec-scale grids rapidly becomes intractable for both a full-simulation CNN correction, as well as a typical U-Net architecture\footnote{For the U-Net architecture, we approximate memory requirements under the assumption of a 5-layer network with channels increasing from 8 to 128 over the layers.}; indeed, for survey volumes up of \(V \approx 8h^{-1}\,\mathrm{Gpc}\), a U-Net style architecture would require 4.6TB of RAM to fit even one sample in memory during training. Instead, we train on smaller \(50^3\) patches (\(\sim 195\,h^{-1}\,\mathrm{Mpc}\)) and then tile our model over the full volume at inference time, which keeps the GPU RAM requirement constant across arbitrary survey volumes during training. Moreover, as demonstrated in Section \ref{sec:scaling_analysis}, this method can then be seamlessly applied to even larger simulation volumes than those used during training, further reducing the overhead required to generate multi-gigapersec scale simulation volumes.

Beyond demonstrating this pipeline for dark matter halos and galaxy catalogs, future work will naturally extend to more realistic survey geometry and observational systematics, such as fiber collisions and survey selection. As data volumes expand, the hybrid approach outlined here—standard reconstruction at large scales plus a trained, subgrid CNN correction—offers a practical, scalable route to sharpen BAO features, recover small-scale information, and ultimately derive tighter constraints on cosmology from next-generation surveys.

\appendix

\acknowledgments

We thank ChangHoon Hahn, Shirley Ho, Francois Lanusse, Chirag Modi, and David Valcin for their helpful discussions and valuable insights. LP acknowledges funding from the NSF GRFP. This work was supported by NSF CDSE grant number AST-2408026.


\bibliographystyle{JHEP}
\bibliography{bibliography.bib}


\end{document}